\documentstyle[graphicx,url]{mn}
\newcommand{\fig}{\includegraphics}
\newcommand{\kmsmpc}{\>{\rm km}\,{\rm s}^{-1}\,{\rm Mpc}^{-1}}
\newcommand{\flux}{\rm {erg\,cm^{-2}\,s^{-1}}}

\newcommand{\kev}{\rm{keV}}
\newcommand{\beq}{\begin{equation}}
\newcommand{\eeq}{\end{equation}}
\def\log{\,{\rm log}\,}

\begin{document}
\title{The soft X-ray properties of quasars in the Sloan Digital Sky Survey}
\author[Shen et al.]
{Shiyin Shen$^{1,2,3}$,  Simon D.M. White$^{2}$, H.J. Mo$^{4}$, Wolfgang
Voges$^{3}$, \and Guinevere Kauffmann$^{2}$, Christy Tremonti$^{5}$, Scott F.
Anderson$^{6}$
\thanks {E-mail: ssy@shao.ac.cn}
 \\
$^1$ Shanghai Astronomical Observatory, Chinese Academy of Sciences, Shanghai
200030, China\\
$^2$ Max-Planck-Institut f\"ur Astrophysik, Karl Schwarzschild
         Str. 1, Postfach 1317, 85741 Garching, Germany\\
$^3$ Max-Planck-Institut f\"ur extraterrestrische Physik, Postfach 1312,
85741 Garching, Germany \\
$4$ Department of Astronomy, University of Massachusetts, Amherst MA
01003-9905, USA\\
$5$ Steward Observatory, 933 N Cherry Ave, Tucson, AZ 85721, USA\\
$6$ Department of Astronomy, University of Washington, Box 351580, Seattle, WA
98195, USA
 }

\maketitle

\begin{abstract}
We use the ROSAT All Sky Survey (RASS) to study the soft X-ray properties of a
homogeneous sample of 46,420 quasars selected from the third data release of
the Sloan Digital Sky Survey (SDSS).  Optical luminosities, both at rest-frame
2500\AA\, ($L_{2500}$) and in [OIII] ($L_{[\rm{OIII}]}$) span more than
three orders of magnitude, while redshifts range over $0.1<z<5.4$. We detect
3366 quasars directly in the observed 0.1--2.4 keV band. Sub-samples of
radio-loud and radio-quiet objects (RLQs and RQQs) are obtained by
cross-matching with the FIRST catalogue. We study the distribution of X-ray
luminosity as a function of optical luminosity, redshift and radio power using
both individual detections and stacks of complete sets of similar quasars. At
every optical luminosity and redshift $\log L_{2\kev}$ is, to a good
approximation, normally distributed with dispersion $\sim 0.40$, at least
brightwards of the median X-ray luminosity. This median X-ray luminosity of
quasars is a power law of optical luminosity with index $\sim 0.53$ for
$L_{2500}$ and $\sim 0.30$ for $L_{[\rm{OIII}]}$.  RLQs are systematically
brighter than RQQs by about a factor of 2 at given optical luminosity. The
zero-points of these relations increase systematically with redshift, possibly
in different ways for RLQs and RQQs.  Evolution is particularly strong at low
redshift and if the optical luminosity is characterised by $L_{[\rm{OIII}]}$.
At low redshift and at given $L_{[\rm{OIII}]}$ the soft X-ray emission from
type II AGN is more than 100 times weaker than that from type I AGN.
\end{abstract}

\begin{keywords}
galaxies: active - galaxies: evolution - galaxies: nuclei - quasars general -
X-rays: galaxies - X-rays: general
\end{keywords}

\section{introduction}

Quasars show strong emission at both ultraviolet (UV) and X-ray wavelengths.
Indeed, many quasar catalogues have been constructed on the basis of their UV
and X-ray properties (e.g. Schneider et al. 2003, 2005; Wolf et al. 2004). The
relation between the UV and X-ray continuum emission is usually characterized
by a spectral index,
 \beq
 \alpha_{OX} \equiv -\frac{\log(L_{2\kev}/L_{2500})}
{\log(\nu_{2\kev}/\nu_{2500})} = -0.384\log(L_{2\kev}/L_{2500})\,,
 \eeq
where $L_{2\kev}$ and $L_{2500}$ are luminosities per unit frequency at
wavelengths of 2\,keV and 2500\AA, respectively.

Previous investigations have indicated that the value of $\alpha_{OX}$ changes
systematically with optical luminosity $L_O$ (e.g. Zamorani et al. 1981; Avni
\& Tananbaum 1982, 1986; Bechtold et al. 1994; Pickering, Impey \& Foltz 1994;
Avni, Worrall \& Morgan 1995; Green et al. 1995; Vignali, Brandt \& Schneider
2003a; Anderson et al. 2004 and references therein). Most of these studies also
concluded that $\alpha_{OX}$ depends only weakly on redshift, but separation of
the redshift and luminosity dependences is difficult because of the strong
correlation between the two quantities in flux-limited samples (e.g. Anderson
\& Margon 1987; Bechtold et al. 2003). A dependence of $\alpha_{OX}$ on $L_O$
implies that the relation between $L_X$ and $L_O$ is nonlinear, the trend in
the observational data being represented by a power law,
 \beq
 L_X\propto L_O^e,
 \eeq
with $e<1$. However, this trend can be affected by a variety of observational
factors, for example, sample definition, sample completeness, observational
errors, and the adopted fitting method. Consequently, the exact form of the
mean $L_X$-$L_O$ relation is still controversial (Franceschini et al.  1994;
Yuan et al. 1998; Vignali et al. 2003b). In particular, as pointed out by La
Franca et al. (1995) and Yuan, Siebert \& Brinkmann (1997), for some earlier
datasets it was conceivable that large photometric errors in the optical
luminosities had biased an intrinsically linear relation into an apparently
nonlinear one.

In order to establish the $L_O$-$L_X$ relation robustly, a large and
homogeneous quasar sample with accurate optical and X-ray measurements is
required. The quasar catalogue selected from the SDSS (York et al. 2000) third
data release (DR3) includes more than 46,000 quasars (Schneider et al. 2005)
and is currently by far the largest available (as of July 2005). The SDSS
quasar selection algorithm is quite efficient, and the completeness at $z<3$ is
at least 90 percent (Richard et al. 2002; Vanden Berk et al. 2005). Deep
surveys which can be used to select X-ray AGN to faint limits cover small areas
of the sky (e.g. Wolf et al. 2004) and so overlap at most a small fraction of
the SDSS data (e.g. Risaliti \& Elvis 2005). On the other hand, the ROSAT
All-Sky Survey (RASS) is quite shallow, with an average exposure time of about
$\sim400$s, and does not detect most SDSS quasars individually.  It is
possible, however, to detect these objects statistically by stacking their
X-ray images. Such stacking is widely used in X-ray astronomy for objects which
are individually below the detection limit (e.g. Wu \& Anderson 1992; Schartel
et al. 1996; Nandra et al. 2002; Georgakakis et al. 2003). Since the number of
SDSS quasars is large, the detection limits can be greatly improved by stacking
many objects with similar optical properties.

Different classes of quasars are observed to have different X-ray properties.
For example, radio-loud objects are systematically brighter in X-rays than
radio-quiet ones (e.g. Ku, Helfand \& Lucy 1980; Zamorani et al. 1981; Bassett
et al. 2004). Type II quasars are usually much weaker in the soft X-ray band
than type I objects, presumably due to stronger absorption (e.g.  Zakamska et
al. 2004; Vignali, Alexander \& Comastri 2004). BL Lacertae (BL Lac) objects
show strong and rapid variability in all bands. In combination, the RASS and
the large quasar catalogue provided by SDSS enable a comparative study of
quasars in these different categories.

In this paper, we use the RASS to study the soft X-ray properties of the SDSS
DR3 quasars. Optical properties for the sample are derived from the SDSS
photometry and spectroscopy. We will make particular use of the continuum
luminosity at 2500\AA\, rest wavelength $L_{2500}$ and the
[OIII]$\lambda$5007 line luminosity $L_{[\rm{OIII}]}$, both of which can be
determined accurately for large samples of quasars. Our analysis considers
X-ray fluxes both for individually detected quasars, and for stacks of quasars
of similar redshift and optical luminosity. We develop statistical methods
specifically designed to study the joint distribution of soft X-ray and optical
properties in the SDSS-RASS samples. In addition, we use the FIRST survey
(Becker, White \& Helfand 1995) to split the SDSS quasars into radio-loud and
radio-quiet classes, and we study the X-ray properties of these two sub-samples
separately. The SDSS quasar catalogue does not include type II and BL Lac
objects (Schneider et al. 2005), but a sample of nearby type II AGN has been
constructed from the SDSS galaxy catalogues by Kauffmann et al.  (2003). We
will use this sample to measure the average soft X-ray luminosities of such
objects.

Our paper is organized as follows. In Section 2 we introduce the SDSS DR3
quasar sample and explain how we define their optical and radio
properties. Section 3 describes our RASS detection technique, both for
individual objects and for stacks. In Section 4, we use a variety of
techniques to study the joint $L_X$ -- $L_O$ distribution. We derive mean
relations from complete sets of X-ray detections of stacks spanning the full
redshift and optical luminosity range of the sample. We use individual
detections to study the high luminosity tail of the distribution of X-ray
luminosity at given optical luminosity. We use complete sets of detections for
stacks of objects which are {\it not} individually detected to study the lower
luminosity part of this distribution.  All approaches give consistent
results. We also split the quasar sample in various ways to examine the
dependence of the relation on redshift, on radio emission strength and on the
presence or absence of broad optical emission lines.  We discuss our
results and draw our conclusions in Section 5.

\section{sample}

\subsection{Optical properties}

The SDSS DR3 quasar catalogue consists of 46,420 objects with luminosities
brighter than $M_i=-22$, with at least one emission line with FWHM (Full Width
at Half Maximum) larger than 1000km~s$^{-1}$ and with highly reliable
redshifts. (Throughout we will assume the standard $\Lambda-$cosmology with
$H_0=70\kmsmpc$, $\Omega_0=0.3$, $\Omega_\Lambda=0.7$.) A few unambiguous broad
absorption line quasars are also included. The sky coverage of the sample is
about 4188 $\rm{deg^2}$ and the redshifts range from 0.08 to 5.41. The
five-band ($u,g,r,i,z$) magnitudes have typical errors of about 0.03 mag. The
spectra cover the wavelength range from 3800 to 9200\AA, with a resolution of
about 1800-2000 (see Schneider et al. 2005 for details).

We use $L_{2500}$, the continuum luminosity at rest wavelength 2500\AA, to
characterize the near-UV luminosities of quasars. We measure the rest-frame
2500\AA\, monochromatic continuum flux, $f_\lambda$(2500\AA), directly from the
SDSS spectra for the quasars with $2500(1+z)$\AA\, in the SDSS spectroscopic
range 3800-9100\AA. For quasars outside this redshift range we assume the shape
of the spectrum to be the same as that of the composite quasar spectrum
presented by Vanden Berk et al. (2001) and we normalize using the rest-frame
continuum flux at 3700\AA\, and 1470\AA\, for low($z<0.5$) and
high($2.7<z<5.25$) redshift quasars respectively. For the few quasars at
$z>5.25$, the rest-frame continuum flux at 1360\AA\, is used to normalize the
spectra. Following Strateva et al. (2005), we use the Schlegel, Finkbeiner \&
Davis(1998) extinction maps to estimate Galactic reddening $E(B-V)$ at the
position of each quasar and the extinction law of Nandy et al. (1975)  to
estimate the Galactic extinction $A_\lambda$ at $2500(1+z)$\AA\, or at the
relevant normalizing wavelength for each quasar. In addition to a strong UV
continuum, most quasars are also characterized by broad emission lines. In this
study, we use the [OIII]$\lambda$5007 line luminosity $L_{[\rm{OIII}]}$ (again
corrected for Galactic extinction) as a measure of the strength of this
emission. Due to the SDSS wavelength coverage, [OIII]$\lambda$5007 can be
measured only for quasars at $z<0.8$. There are 9103 such objects.

\begin{figure}
\fig[width=0.48\textwidth]{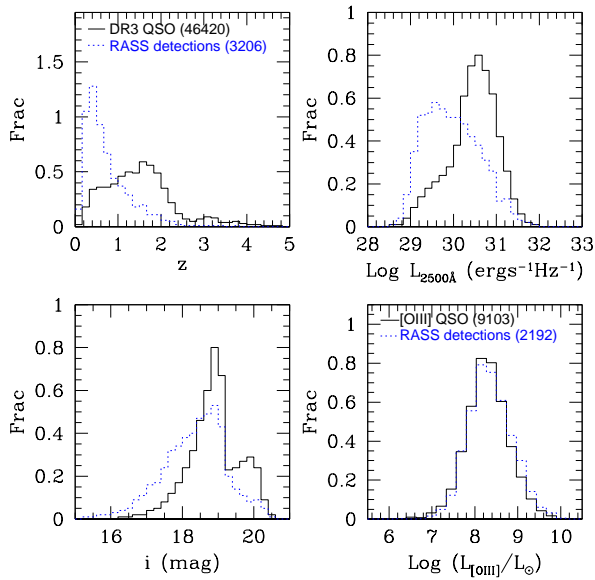} \caption[histogram]{The distributions of
quasars in the SDSS DR3 catalog with respect to some basic parameters: redshift
(upper left), rest-frame 2500\AA\, luminosity (upper right), $i-$band apparent
magnitude (lower left), and [OIII] line luminosity (lower right). The solid
lines show results for the full sample whereas the dotted lines show the
results for quasars with individual RASS detections (see Section \ref{XX}). All
the histograms are normalised to unit total area.}
 \label{hBasic}
\end{figure}

To illustrate the general properties of the SDSS DR3 quasar sample, we show in
Fig. \ref{hBasic} their distributions in redshift $z$, in luminosity
$L_{2500}$, and in $i$-band apparent magnitude. The drop at $i\sim19.1$
roughly corresponds to the completeness limit of the low redshift ($z<3$)
multi-colour selected sample (Richards et al. 2002). The $L_{[\rm{OIII}]}$
distribution of the low-redshift objects is shown in the lower right panel.
These plots also show the distribution for quasars detected individually in the
RASS. These objects are clearly biased towards low redshifts, low continuum
luminosities and bright apparent magnitudes. Interestingly, there is no
apparent bias in the line luminosity distribution.

\subsection{Radio properties}

The X-ray properties of radio-loud quasars (RLQs) and radio-quiet
quasars(RQQs) are significantly different. Compared with RQQs, RLQs are
characterized by higher X-ray luminosities and flatter X-ray spectra at given
optical luminosity (e.g. Worrall et al. 1987; Green et al. 1995; Schartel et
al. 1996; Brinkmann et al. 2000; Bassett et al. 2004). However, RLQs are a
small minority of the quasar population, only about ten percent of the total
(Brinkmann et al. 2000; Ivezic et al. 2002).

\begin{figure}
\fig[width=0.48\textwidth]{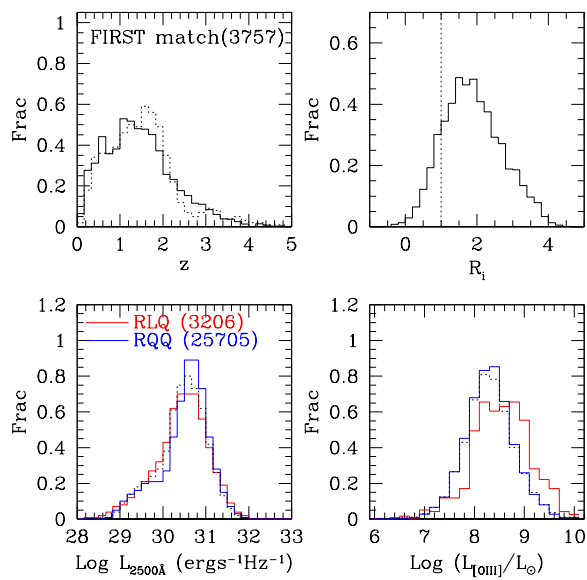} \caption[histogram]{Our sub-samples of
radio-loud and radio-quiet quasars. The top two panels show the $z$ and $R_i$
distributions of DR3 quasars with a FIRST catalogue match. The bottom two
panels show histograms of the $L_{2500}$ and $L_{[\rm{OIII}]}$ distributions
for radio-loud (RLQ; red) and radio-quiet (RQQ; blue) quasars. For comparison,
the $z$, $L_{2500}$ and $L_{[\rm{OIII}]}$ distributions for the full sample
of DR3 quasars are shown by dotted lines. All the histograms are normalised to
unity.}
 \label{RLRQ}
\end{figure}

In the DR3 quasar catalogue, there are 3757 objects having FIRST matches within
the matching radius 2.0 arcsec.  We show the redshift distribution of these
objects in the top left panel of Fig. \ref{RLRQ}. In the sky covered by both
the SDSS DR3 and FIRST, there are 37980 quasars without a FIRST match. Unlike
the RASS quasars, the $z$ distribution of the FIRST quasars is almost the same
as that for the general catalogue (dotted line), which suggests that the
optical-to-radio spectral index does not depend strongly on redshift or optical
luminosity.

Ivezic et al. (2002) have investigated the radio properties of SDSS objects in
detail and suggest a bimodal distribution of radio-to-optical flux ratio,
$R_i$, which separates RQQs from RLQs at $R_i\sim 1$ . Here $R_i$ is defined as

 \begin{eqnarray}
 R_i=\log(F_{radio}/F_{optical})
 =0.4(m_i-t)
 \nonumber\\
 ~~~ {\rm{with}} ~~~
t=-2.5\log\left(\frac{F_{int}}{3631\,\rm{Jy}}\right)\,,
 \end{eqnarray}
where $m_i$ is the $i$-band magnitude, $t$ is the AB radio magnitude, and
$F_{int}$ is the integrated 20 cm continuum flux density listed in the FIRST
catalogue. We show the $R_i$ distribution of the 3757 FIRST quasars  in the top
right  panel of Fig. \ref{RLRQ}.

We select quasars with $R_i>1$ as RLQs, giving a sample of 3206 objects. Since
the FIRST sensitivity limit of 1mJy corresponds to $t\approx16.4$ (Ivezic et
al. 2002), bright quasars with $i<19.1$ within the FIRST sky area but with no
FIRST catalogue match have $R_i<1.08$. We select such objects as RQQs, giving
a sample of 25705. Apart from the 3206 RLQs and 25705 RQQs, there are a
further 17509 objects with indeterminate radio properties. When studying
quasars without specifying their radio properties, we will use the full sample
of 46420 objects. We show the $L_{2500}$ and $L_{[\rm{OIII}]}$
distributions of RLQs and RQQs as the blue and red histograms in the lower two
panels of Fig.  \ref{RLRQ}. As we can see, although RQQs are selected with a
bright magnitude limit ($i<19.1$) their $L_{2500}$ distribution shows no
clear difference from that of the RLQs. However, the $L_{[\rm{OIII}]}$
distribution of RLQs is biased somewhat high in comparison with that of
RQQs. RLQs tend to have slightly higher equivalent widths of
[OIII]$\lambda$5007 than RQQs (e.g.  Marziani et al. 2003 ). The numbers of
quasars in the RLQ, RQQ and full samples are listed together with the relevant
magnitude limits and redshift ranges in Table \ref{sample}.

\section{The X-ray data}

\subsection{Individual X-ray detections}\label{XX}

The RASS mapped the sky in the soft X-ray band (0.1--2.4\,\kev) with an
effective exposure time varying between 400 and 40,000s, depending on
direction. The angular resolution of the survey is 0.5 arcmin and the total
exposure time is $\sim 1.0\times10^7$ s. Two source catalogues were generated
based on RASS images: the Bright Source Catalogue (BSC, Voges et al. 1999) and
the Faint Source Catalogue (FSC, Voges et al. 2000). The BSC contains $\sim$
18,800 objects with detection likelihood $L$ [defined after equ. (\ref{P})
below] of at least 15, while the FSC includes $\sim$105,900 objects with detection
likelihood between 7 and 15. In the DR3 quasar catalogue, the position of each
quasar was matched to these two catalogues out to a radius of 30 arcsec. The
number of matches is 2672. The contamination from chance superpositions is
estimated to be about 1--2 percent. However, since the position of each SDSS
quasar is known very well, to better than one acrsec according to Pier et al.
(2003), we applied the upper-limit maximum likelihood method (the COMP/UPP
command in the MIDAS/EXSAS package) directly to the RASS images to calculate a
detection likelihood for each SDSS quasar. To be consistent with the RASS
source catalogues, we accept all sources with detection likelihood greater than
or equal to 7 as individual X-ray detections.

Our detection algorithm is applied to the RASS-II data, which includes 1,378
sky fields, each having a size of $6.4^\circ \times 6.4^\circ$, and with
neighboring fields overlapping at least 0.23 degrees (Voges et al. 1999). The
data in each field is binned into a $512\times512$ pixel image, each pixel
having a size of 45 arcsec in the 0.1--2.4\,keV band. A local-detection method
is first used to search for source candidates in the map within windows of
increasing size, starting from $3\times3$ pixels. Identified source candidates
are then cut out of the image and a smooth background is built by spline
fitting. Using the background and exposure map of each field, the upper-limit
maximum likelihood detection method takes into account the position of each
photon relative to the position of each source candidate within the extraction
radius and so accumulates a source detection likelihood $L$. All photons in the
0.1--2.4\,keV band are used and the extraction radius is chosen to be 5 times
the FWHM (60 arcsec in RASS) of the point spread function (PSF). With this
algorithm, 3366 quasars are detected with $L\geq 7$, which is 26\% higher than
the number of RASS catalogue matches. This number must still include
some contamination from random fluctuations in the background.

To estimate the extent of contamination, we build a mock catalogue with the
same number of objects and the same sky coverage as the SDSS DR3 quasar sample.
The mock objects have random sky positions except that they avoid disks of
radius 30 arcsec centered at the position of each DR3 quasar. Applying our
algorithm to this mock catalogue, we get 246 random positions with detection
likelihood $L\geq 7$, suggesting that a similar number of our quasar
``detections'' are spurious.  To account for this contamination statistically,
we randomly select 246 quasars from our 3366 individual detections, requiring
not only the same X-ray flux distribution as the random positions with RASS
``detections'', but also $z$ and $L_{2500}$ values chosen randomly from
those of the entire DR3 sample of 46420 quasars. These 246 quasars are then
marked as possible contaminants and are excluded from further analysis. The
number of RASS individual detections in the sample we finally analyze is
therefore 3120. Of course, there is no reason to identify any particular quasar
as spurious. Our exclusion of 246 objects is only intended to be statistically
correct.

We showed the distributions of the basic properties of the quasars with RASS
individual detections as dotted lines in Fig. \ref{hBasic}. Compared with the
general SDSS population, these objects are biased towards low redshifts
and bright apparent magnitudes.

Among the 3206 RLQs, 379 have RASS individual detections, and the estimated
number of contaminants is 36. For the 25705 RQQs, the number of detections and
estimated number of contaminants are 2164 and 130, respectively. These
numbers are also listed in Table \ref{sample}. The individual detection
fraction is significantly higher for RLQs than for RQQs. As we will show, this
is because the RLQs have systematically stronger X-ray emission.

\begin{figure}
\fig[width=0.48\textwidth]{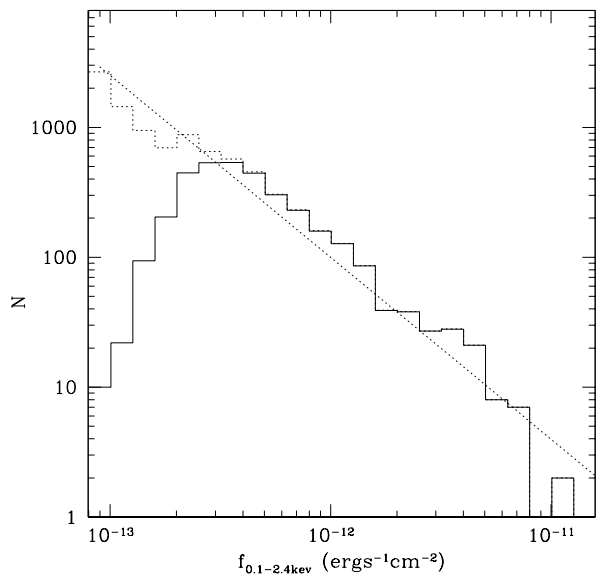} \caption[histogram of flux]{ Histograms of
X-ray flux for our contamination-corrected sample of individually detected DR3
quasars in the 0.1--2.4\,keV band. The solid histogram shows the raw data
whereas the dotted line indicates the effect of correcting for incompleteness
as described in Section \ref{comp}. A dotted straight line with log N--log S slope
of $-1.4$ is shown for comparsion.}
 \label{hflux}
\end{figure}

We convert the count rates of quasars with individual RASS detections to fluxes
in the 0.1--2.4\,keV band using PIMMS\footnote{\url{
http://heasarc.gsfc.nasa.gov/docs/software/tools/pimms_install.html}}.  Here we
assume the photon energy distribution to be a simple power law, $N(E)\propto
E^{-\Gamma}$ with $\Gamma=2$, and we correct for an absorbing column fixed at
the Galactic value for each line-of-sight (Dickey \& Lockman 1990). Spectral
studies of quasars in the 2--10\,keV band give a photon index $\Gamma\sim 1.9$
for radio quiet objects and flatter spectra, $\Gamma \sim 1.6$, for radio-loud
objects (e.g. Reeves \& Turner 2000). In the soft X-ray band, quasar spectra
are known to be steeper but with significant scatter (e.g.  Yuan et al. 1997;
Brinkmann et al. 1997, 2000). Although the values of $\Gamma$ are quite
uncertain, the assumption of a flatter ($\Gamma\sim 1.5$) or a steeper
($\Gamma\sim 2.5$) spectrum only changes the derived flux by a few percent. The
flux distribution of the 3120 individual RASS detections in our
contamination-corrected sample is shown as a solid histogram in Fig.
\ref{hflux}. The peak of the flux distribution is at about
$4\times10^{-13}\flux$, which roughly corresponds to the flux completeness
limit of RASS.  We discuss the X-ray completeness issue in detail in the next
section.

\subsection{Detection completeness}\label{comp}

In this section, we consider the completeness of our catalogue of individual
RASS detections. Our aim is to figure out the limit above which our sample is
complete, and the correction for incompleteness which is needed below this
limit.

Whether or not a source of given X-ray flux is detected in the RASS depends
on factors such as the effective exposure time, Galactic absorption and
background noise. Thus, completeness depends on location in the sky.  Since
the selection of SDSS quasars is independent of the RASS parameters and the
number of these quasars is quite large, their positions can be viewed as a
good random realization of the RASS sky inside the DR3 area.  The RASS
detection completeness $C_{\rm RASS}$ at a given X-ray flux $f$ can thus be
defined as \footnote{Considering the possible variance of the RASS exposure
time and Galactic hydrogen column density over large scales, it is important
to note that $C_{\rm RASS}$ as defined here applies only to the sky area of
SDSS DR3.}
\beq\label{C}
C_{\rm RASS}(f)=\sum_{i=1}^{46420} D_i(f) /46420\, ,
\eeq
where $D_i$ equals $1$ if a source with flux $f$ can be detected by RASS at
the position of the $i$th quasar, and $D_i=0$ otherwise.

More specifically, for an assumed source with flux $f$ at a given position, we
calculate the predicted count-rate assuming a photon index $\Gamma=2$ and the
Galactic hydrogen absorption column. The expected number of  photons $n_s$ is
then the product of the effective exposure time at that position and the
count-rate. We compare $n_s$ with the local background and define the detection
probability $P$ as
 \beq\label{P}
  P=\sum_{n=0}^{n_s+n_b-1} \frac{{n_b}^n}{n!} e^{-n_b}\,
 \eeq
where $n_b$ is the expected number of background photons inside a circle of 90
arcsec (1.5 times the FWHM, about 3$\sigma$ for a Gaussian PSF) surrounding the
source position. The simple Poisson counting statistics assumed in equation
(\ref{P}) do not work properly when the number of background photons is very
small. For example, a detection with $n_s=1$ can easily satisfy the condition
$L\geq7$ [$L$ is defined as $L =-\ln (1-P)$] if $n_b$ is very small. We
therefore introduce another criterion for a positive source detection,
 \beq\label{Pcond}
  n_s \geq 2.5+(10n_b)^{1/2}\,
 \eeq
which is derived from the maximum likelihood correction for
the Eddington Bias (Wang 2004). Objects satisfying both equation
(\ref{Pcond}) and $L \geq 7$ are accepted as individual
detections, for which $D_i=1$. For all other objects, $D_i=0$.

\begin{figure}
 \centering
\fig[width=0.48\textwidth]{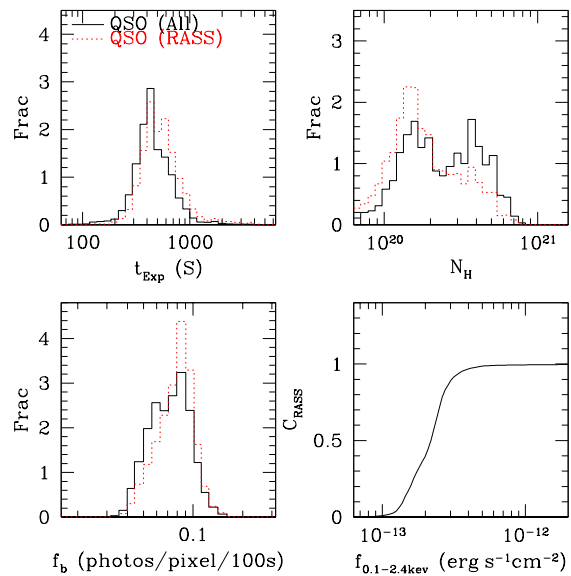} \caption{The distributions of RASS
exposure time, Galactic hydrogen column density and RASS background level at
the positions of SDSS quasars (upper left, upper right and lower left
respectively). The solid lines show the results for  all SDSS DR3 quasars,
whereas the dotted lines represent  quasars with  individual RASS detections.
The lower right panel show the RASS detection completeness as a function of
X-ray flux. The histograms are normalised to unit enclosed area.}
 \label{CRASS}
\end{figure}

We show  histograms of the Galactic hydrogen column, the RASS effective
exposure time and the background level at the position of  the 46420 SDSS
quasars in Fig. \ref{CRASS}. The background level is expressed as the expected
number of background photons per pixel for an  exposure time of 100 seconds.
Distributions for objects with individual detections are also shown for
comparison. As one can see, the individually detected objects  are biased to
longer exposure times and lower Galactic absorption column. The distribution of
the background level is slightly biased to higher values  because of the
anti-correlation between Galactic hydrogen density and background level. Based
on the distribution of the above three quantities, we derive the completeness
of our RASS quasar detections as a function of X-ray flux in the 0.1--2.4\,keV
band from equations (\ref{C}), (\ref{P}) and (\ref{Pcond}). The results are
shown in the lower right-hand panel of Fig. \ref{CRASS}. Our RASS individual
detections are complete to $\sim$ 98 percent for
$f_{0.1-2.4\rm{keV}}>5.0\times10^{-13}\flux$. The lower flux limit
$f_{\rm{lim}}$, at which $C_{\rm RASS}\sim 0$, is at $\sim
8.0\times10^{-14}\flux$, consistent with the fact that 3365 of 3366  RASS
individual detections have flux greater than $f_{\rm{lim}}$. Using $C_{\rm
RASS}$, we can correct for the incompleteness of our RASS individual detections
and estimate the true number of quasars at  given X-ray flux above
$f_{\rm{lim}}$. From this, we estimate $\sim 9397$ objects with fluxes greater
than $8.0\times10^{-14}\flux$, as shown in the dotted histogram in Fig.
\ref{hflux}. As we can see, the slope of the $\log N$-$\log S$ distribution
after correction is close to $-1.4$.

\subsection{Detection of stacks}\label{Xstack}

As discussed above, only about 7 percent of the SDSS quasars have individual
RASS detections because of the limited exposure time. The X-ray properties of
the detected quasars are likely to be biased relative to those of the sample as
a whole. We have therefore developed a stacking algorithm to study the X-ray
properties of the full sample in a way which allows such biases to be
understood and corrected.

In general, if the images of $N$ objects are stacked and if we assume these
objects all have similar X-ray properties, then the number of photons in the
signal increases by a factor of $N$, while the noise from the background
fluctuation increases only by a factor of $\sqrt{N}$. Therefore, the
signal-to-noise ratio of the stacked image (the stack) is about $\sqrt{N}$
times higher than that of a single image. To proceed, we take the original
($512\times 512$ pixel) images and exposure maps for the RASS-II fields and
cut out small binned images and exposure maps of $65\times65$ pixels, centered
on the sources to be stacked. Since neighboring RASS fields overlap each other
by about 0.23 degrees (18 pixels), a few (less than 1 percent) of the SDSS
objects are too close to the image boundaries to make these small images. The
small images and the corresponding exposure maps are then added to make the
stack. We apply our local detection criterion to the stack using spline
fitting to get a smooth background. The photon-event tables of the
corresponding binned images are also merged into one table, with the
coordinates of all the quasars shifted to be the same. The detector pixel
coordinates of all the photon events are kept so as to account for vignetting
in the final stack image. With the merged photon table, the stacked exposure
map and the background from the stacked image, we apply the same detection
method as used for single sources.

 For a stack of $N$ sources with Galactic hydrogen density $N_{H,i}$, redshift
$z_i$, RASS effective exposure time $t_i$ and X-ray luminosity
$L_{2{\kev},i}$, the number of source photons $N_s$ is \beq \label{stack}
N_s=\sum_{i=1}^N{L_{2{\kev},i}~g(N_{H,i}, z_i)t_i}\,, \eeq where $g(N_{H,i},
z_i)$ is a function which converts the rest-frame X-ray luminosity at 2~keV to
observed count-rate. Since the X-ray luminosities of the sources
in the stack are supposed to be similar, the weighted average X-ray luminosity
of a stack at rest-frame 2\,keV is
\beq
\label{Lstack}
L_{S,2\kev}=\frac{N_s}{\sum_{i=1}^N{g(N_{H,i}, z_i)t_i}}.
\eeq
Only stacks with $L \geq 7$ are accepted as detections. For those with $L<7$
the photon counts are used to derive upper limits.

\section{The soft X-ray properties of SDSS quasars}

In this section, we investigate how the X-ray properties of SDSS quasars
depend on optical luminosity, on redshift and on [OIII] line luminosity, both
for our sample as a whole and for our subsamples of radio-loud and radio-quiet
quasars. We first present results based on stacks which group all objects with
similar optical luminosity and redshift. We then develop a maximum likelihood
method which allows individual detections to be used to study the distribution
of $L_{2\kev}$ at fixed optical luminosity and redshift. Based on the results,
we predict the average $L_{2\kev}$ of quasars {\it without} individual
detections and compare directly with the values measured from stacks of such
objects.

\subsection{Stacks as a function of $L_{2500}$ and $z$}\label{stackall}

For a first analysis of our data we consider the simple question: what is the
mean $L_{2\kev}$ of QSOs of given $L_{2500}$ and how does it depend on
redshift? To address this, we stack {\it all} the sources in each of our chosen
luminosity and redshift bins, regardless of whether they are individually
detected or not, and we estimate an average X-ray luminosity for the stack.
All our quasars are spectroscopically confirmed, so contamination by spurious
detections is not an issue for this analysis and we do not exclude the
statistically identified ``contaminants'' from our stacks, which we refer to as
``total'' stacks in order to distinguish them from the stacks of non-detected
sources analysed in Section \ref{stacks} below.

\begin{figure}
\centering \fig[width=0.48\textwidth]{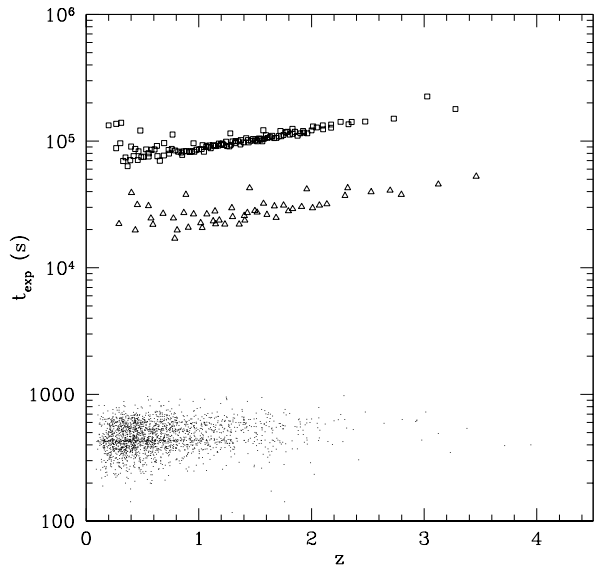} \caption{Exposure time as a
function of redshift for our stacks and for individual sources. The dots show
quasars with individual detections whereas the triangles and squares refer to
stacks of RLQs and of RQQs respectively. } \label{sobj}
\end{figure}

We first divide the sources into seven $\log L_{2500}$ bins from 28.5 to
32.0, with a width of 0.5. The objects in each luminosity bin are then further
divided into redshift bins according to the total summed exposure time. Since
the X-ray signals from higher redshift quasars are weaker (see Fig.
\ref{hBasic}), we stack more objects (i.e. require a longer total exposure
time) for higher redshift bins. This strategy is illustrated in Fig.
\ref{sobj}, where the total exposure time of the stacks of RLQs (triangles) and
RQQs (squares) are plotted against their average redshifts. Since the average
X-ray luminosity is expected to be weaker for RQQs than for RLQs, we group more
objects into one stack in the radio-quiet subsample. We finally get 51 stacks
of RLQs and 118 stacks of RQQs.

For the RLQs, we detect 46 stacks with $L>7$ based on counts in the full
0.1--2.4 keV band but 49 based on counts in the restricted 0.5--2.0 keV band.
For RQQs, the numbers of detections in the full and restricted bands are 107
and 117 respectively. The higher detection efficiency in the hard band
(0.5--2.0keV) is a result of the weaker background there (Snowden et al. 1995).
To investigate this further we assume that the photon index $\Gamma$ can be
taken as constant for all RQQ stacks and for all RLQ stacks, but may differ
between the two classes. Using equation (\ref{Lstack}), we then estimate the
average $L_{2\kev}$ of each stack twice, once from the photon counts in
0.1--2.4 keV and once from the counts in 0.5--2.0 keV.  We denote the X-ray
luminosity estimated from the 0.5--2.0 keV counts by $L_{2\kev,B}$ to
distinguish it from $L_{2\kev}$ estimated from the 0.1--2.4 keV counts. By
requiring the consistency between $L_{2\kev,B}$ and $L_{2\kev}$, we find that
$\Gamma\approx 1.9$ is required for RLQs and $\Gamma \approx 2.1$ for RQQs.
This result agrees with the fact that the RLQs show somewhat flatter X-ray
spectra than RQQs (e.g. e.g. Worrall et al. 1987; Green et al. 1995; Schartel
et al. 1996; Brinkmann et al. 2000; Bassett et al.  2004). With this in mind,
in Section \ref{indiv}, we will calculate $L_{2\kev}$ for individual detected
objects assuming $\Gamma=1.9$ for RLQs and $\Gamma=2.1$ for RQQs.  We note that
in Section \ref{XX}, we used the $\Gamma=2.0$ to convert the photon counts to
fluxes regardless of whether objects are RLQs or RQQs. This slight
inconsistency is too small to affect any of our conclusions.

\begin{figure}
\centering \fig[width=0.48\textwidth]{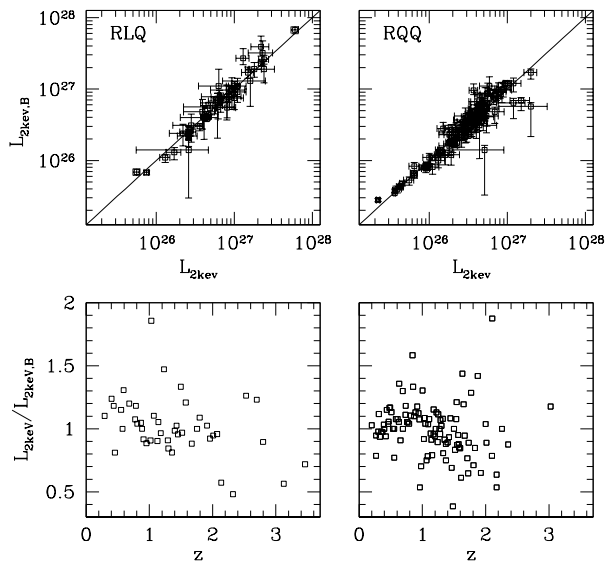} \caption{Comparison of
the average luminosity of stacks at 2 keV estimated from counts in the full
0.1--2.4 keV band ($L_{2\kev}$) with that estimated from counts in the
restricted 0.5--2.0 keV band ($L_{2\kev,B}$). The left two panels show the
results for RLQs and the right two for RQQs. The upper panels show a direct
comparison between $L_{2\kev}$ and $L_{2\kev,B}$, while the lower panels show
the ratio of $L_{2\kev}/L_{2\kev,B}$ as function of the average redshift of the
stacks. The solid lines represent the relation $L_{2\kev}=L_{2\kev,B}$. }
\label{Compare}
\end{figure}

In Fig. \ref{Compare} we compare $L_{2\kev}$ with $L_{2\kev,B}$ for stacks
detected in both bands.  The upper panels demonstrate that the two estimates
agree to well within their errors both for RLQs and for RQQs. In the lower
panels, we plot the ratio $L_{2\kev}/L_{2\kev,B}$ as a function of the average
redshift of the stacks. As we can see, the typical difference is 10 or 20\%
with very weak dependence on redshift. (Linear fitting of
$L_{2\kev}/L_{2\kev,B}$ as a function of redshift gives a slope of $-0.12\pm
0.04$ for RLQs and $0.07\pm0.07$ for RQQs. To be consistent with our individual
detections, we base our stack analysis as far as possible on $L_{2\kev}$
estimated from the 0.1--2.4 keV band.  For the few stacks detected only in the
0.5--2.0 keV band, we use $L_{2\kev,B}$ instead. Table \ref{sample} lists
the final number of stacks detected in each of our samples.

\begin{figure}
\centering \fig[width=0.48\textwidth]{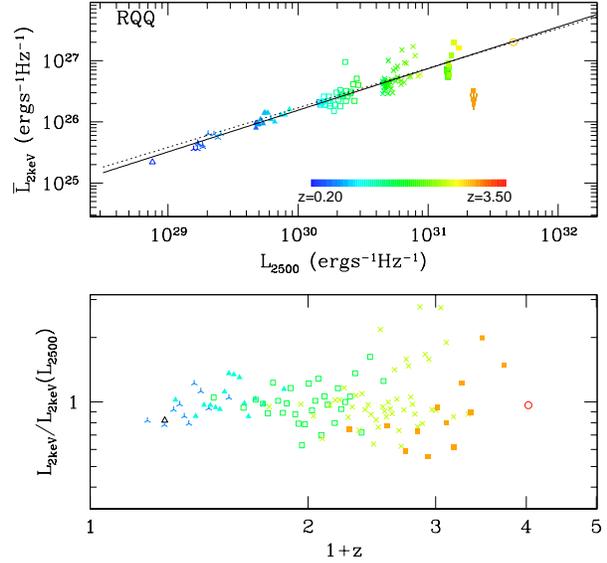} \caption {Mean X-ray
luminosity $L_{2\kev}$ as a function of optical luminosity $L_{2500}$ for RQQs.
Symbols in the top row show the mean $L_{2\kev}$ of our stacks of {\it all}
quasars in each redshift and $L_{2500}$ bin, colour-coded according to mean
quasar redshift. Solid line is the fitting relation of equation (\ref{ARQQ}).
For comparison, the dotted line shows the mean relation between X-ray and
optical luminosity given by Strateva et al. (2005).  In the bottom row we
divide each stack's mean value of $L_{2\kev}$ by the value predicted for its
mean $L_{2500}$ by equation (\ref{ARQQ}) and we plot the result against mean
redshift. Differing symbols distinguish stacks in our 7 different ranges of
$L_{2500}$. Colour-coding here helps to distinguish between symbols of
different type. } \label{All_RQQ}
\end{figure}

\begin{figure}
\centering \fig[width=0.48\textwidth]{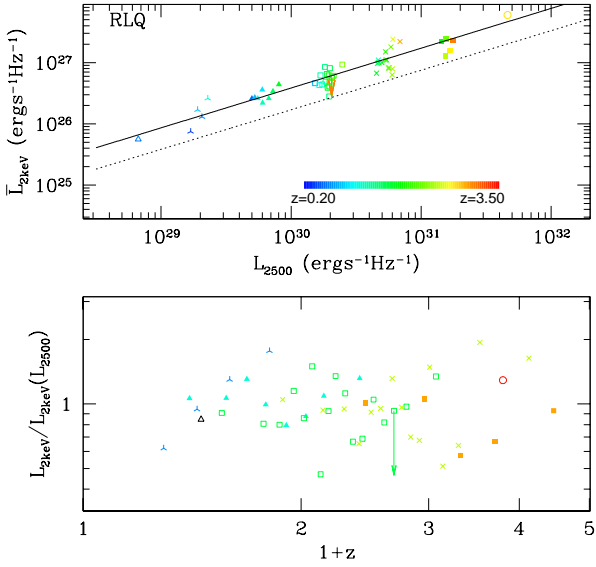} \caption {Mean X-ray
luminosity $L_{2\kev}$ as a function of mean optical luminosity $L_{2500}$
and redshift for stacks of RLQs. The structure of the figure is the same as
that of Fig.  \ref{All_RQQ}. The dotted line again gives the relation
of Strateva et al. (2005).} \label{All_RLQ}
\end{figure}

The upper panels of Fig. \ref{All_RQQ} and Fig. \ref{All_RLQ} show the average
$L_{2\kev}$ of the RQQ and RLQ stacks as a function of their mean $L_{2500}$.
Different symbols refer to our different $L_{2500}$ bins and are colour-coded
according to redshift. Arrows denote 1-$\sigma$ upper limits for the few stacks
we do not detect. Clearly the mean $L_{2\kev}$ values of the stacks correlate
extremely well with their mean $L_{2500}$ and the relation between them is well
represented by a power law. We fit a linear relation to this plot
\beq\label{lfit} \log \bar{L}_{2\kev} = \alpha(\log L_{2500}-30.5) + \delta
\eeq by minimizing $\chi^2$ weighting each point by an ``error'' which is the
sum in quadrature of two terms, one coming from the observational error in the
mean $L_{2\kev}$ measurement for each stack, the other from the expected
population variance in the mean for a stack of given size. As we will show in
section \ref{cdis}, the $L_{2\kev}$ values of quasars at given $L_{2500}$ and
$z$ follow a log-normal distribution with typical scatter $\sim 0.40$. Since
each of our stacks contains hundreds of sources, the population variance term
is smaller than that due to the observational errors. The best-fit linear
relations are
 \beq \label{ARQQ}
\log \bar{L}_{2\kev} = (0.68\pm0.01)(\log L_{2500}-30.5)+ (26.53\pm0.01)
 \eeq
for RQQs and
 \beq\label{ARLQ}
\log \bar{L}_{2\kev} = (0.65^{+0.02}_{-0.03})(\log L_{2500}-30.5) + (26.91^{+0.02}_{-0.01})
 \eeq
for RLQs. These are shown as solid lines in Fig.
\ref{All_RQQ} and Fig. \ref{All_RLQ} respectively. We also bin the full sample
of QSOs into 209 $L_{2500}$ and redshift bins without regard to radio
properties, detecting 199 of them with $L>7$.  The best power-law relation
between $\bar{L}_{2\kev}$ and $L_{2500}$ is
 \beq \label{All}
\log\bar{L}_{2\kev} = (0.64\pm0.01)(\log L_{2500}-30.5) + (26.63\pm0.01)\,.
 \eeq
The result for the full sample is thus similar to that for the RQQs, as expected given the
relatively small fraction of RLQs in the sample. The uncertainties in the above
fits and the corresponding minimum $\chi^2$ values are listed in Table
\ref{table_xi}.

The relations for RLQs and RQQs have similar slope but the mean X-ray
luminosity of RLQs at given $L_{2500}$ is about twice that of RQQs. In the
upper panel of Fig. \ref{All_RQQ}, we have over-plotted the mean relation which
Strateva et al. (2005) obtained for their complete sample of 228 radio-quiet,
non-BAL SDSS AGN with medium-deep ROSAT imaging. More than 80\% of these
sources have individual X-ray detections.  Clearly their results agree well
with what we find here for a 100 times larger sample. However, this dramatic
increase in sample size (which includes an increase in the number of individual
detections by a factor of more than 10) will allow us to make much more precise
statements about the $L_{2\kev}$ -- $L_{2500}$ distribution than was
possible with a sample of 228 objects.

Already in these plots there is some indication for a redshift dependence,
especially for the two lowest $L_{2500}$ bins. This is shown more
explicitly in the lower panels in Fig. \ref{All_RQQ} and Fig. \ref{All_RLQ}
where we have divided the mean $L_{2\kev}$ of each stack by the value
predicted by the above best-fit relations [equations (\ref{ARQQ}) and
(\ref{ARLQ})] applied to the mean $L_{2500}$ for the stack. We plot the
result against the mean redshift of the stack. Although there is no global
trend with redshift for either subsample, trends are visible in both cases
when one compares symbols of the same type, thus stacks with similar mean
$L_{2500}$. These trends are most obvious at low redshift and are stronger
for RLQs than for RQQs.

To test explicitly for the significance of this redshift dependence we assume
mean X-ray luminosity to vary with redshift as
\beq\label{zL3p}
\log L_{2\kev}=\alpha (\log L_{2500} -30.5) + \beta \log(\frac{1+z}{1.5}) +
\delta\,.
\eeq

The three model parameters $\alpha, \beta$ and $\delta$ can then be estimated
by minimizing $\chi^2$ for our sets of stacks. The results, together with the
corresponding minimum $\chi^2$ values, are listed in Table \ref{table_xi}.
Comparing with the values obtained when forcing $\beta=0$ [i.e. fitting to
equation (\ref{lfit})], we see that including a power-law redshift dependence
has substantially improved the quality of the fit for the full sample and for
the sample of RQQs. For RLQs, however, including a redshift dependence of this
form produces little improvement, even though some redshift dependence is
evident in the bottom panel of Fig. \ref{All_RLQ}. This is in part because
$\chi^2$ per degree of freedom is well below unity for both models for this
sample, and in part because the apparent dependence on redshift is poorly
described by a single power law.  Below we use a more complex model which
better describes the apparent behaviour, showing that this leads to a further
significant improvement in $\chi^2$ for our stacks of objects in the total and
RQQ samples, as well as describing well the joint $(L_{2\kev}$ -- $L_{2500},z)$
distribution of individual detections in all three samples.

\subsection{Results from individual detections alone}
\label{indiv}

The previous section presented results for our subsamples of RLQs and RQQs
using stacked data only. Here we consider what can be learned, without using
stacks, from our sample of individually detected sources. As mentioned above,
we calculate the $L_{2\kev}$ of the individual detected QSOs from their photon
count using photon indices $\Gamma=1.9$ for RLQs and $\Gamma=2.1$ for RQQs.
When we consider the full sample of QSOs without regard to radio properties, we
use $\Gamma=2$. Among our 3206 RLQs there are 341 individual detections after
statistical rejection of 36 ``contaminants''. Among our 25689 RQQs there are
2163 detections left after rejection of 130 ``contaminants''. When the full
sample of 46420 DR3 quasars is analysed there are 3120 detections after
rejection of 246 ``contaminants''. These detection statistics are listed for
each sample in Table \ref{sample}.

\subsubsection{The redshift dependence}\label{depz}

In Section \ref{stackall}, we showed evidence that the average X-ray luminosity
of quasars of given $L_{2500}$ depends on redshift both for RLQs and for RQQs.
In this section, we use data from our individual detections to further check
and quantify this dependence.

In the bottom panels of Fig. \ref{All_RQQ} and \ref{All_RLQ}, the $L_{2\kev}$
dependence on redshift is more evident at low redshifts than that at high for
both classes of quasar. Based on this impression, we parameterize the average
$L_{2\kev}$ of quasars of given $L_{2500}$ and $z$ through

 \begin{small}
 \begin{eqnarray}\label{zL5p}
\log L_{2\kev}  =
 \nonumber \\
 \left\{
 \begin{array}{ll}
    \alpha (\log L_{2500} -30.5) + \beta \log(\frac{1+z}{1+z_0}) + \delta
     & \mbox {for ($z<z_0$)}\\
    \alpha (\log L_{2500}-30.5) + \gamma \log(\frac{1+z}{1+z_0}) + \delta
      & \mbox {for ($z>z_0$)}
 \end{array}\right.
\end{eqnarray}
\end{small}
With this assumption, the average $L_{2\kev}$ is proportional to $(1+z)^\beta$
at $z<z_0$ and  proportional to  $(1+z)^\gamma$ at $z>z_0$ but is proportional
to $L_{2500}^\alpha$ at all redshifts. For
$\beta\approx\gamma$, this relation is equivalent to equation (\ref{zL3p}).

As we will verify in Section \ref{cdis}, the distribution of $L_{2\kev}$ at
given $L_{2500}$ and $z$ is well approximated by a log-normal with scatter
$\sim 0.40$, almost independent of $L_{2500}$ and $z$. Adopting this model,
the likelihood that the $i$th SDSS quasar will be detected with X-ray
luminosity $L_{2\kev,i}$ is \beq P_i=f(L_{2\kev,i}|L_{2500,i},z_i) \eeq where
$f$ is the log-normal density function with median given in terms of
$L_{2500}$ and $z$ by equation (\ref{zL5p}) and with scatter $\sim0.40$ [see
equation (\ref{lognormal})]. As shown in Section \ref{comp}, we can calculate
the minimum detectable X-ray flux for a QSO from its position, based on local
values of Galactic hydrogen column, RASS effective exposure time and RASS
background level.  Given the quasar's redshift, we can convert this minimum
flux to a minimum detectable X-ray luminosity $L_{\rm {min}}$. Thus for each
undetected DR3 quasar, we know $L_{2\kev} < L_{\rm {min}}$.  The likelihood of
each non-detections can then be defined as
 \beq
  P_{non,j}=\int_{-\infty}^{L_{\rm{min},j}} f(L_{2\kev}|L_{2500,j},z_j)
 dL_{2\kev}.
 \eeq
For each set of parameters $\alpha, \beta, \gamma, \delta, z_0$ in equation
(\ref{zL5p}), the joint likelihood of $N$ individual detections and $M$
non-detections can be written as
 \beq
 {\mathcal{L}}\propto \prod_{i=1,N}P_i\prod_{j=1,M} P_{non,j}
 \eeq
The 5 model parameters then can be estimated by maximizing $\mathcal{L}$. For
sufficiently large samples (such as those used here) the quantity $2\ln
({\mathcal{L}_{\rm max}}/\mathcal{L})$ is distributed like $\chi^2$ with 3
degrees of freedom.  By marginalizing over some of the parameters we can
construct confidence regions for the other parameters in the standard way. For
comparison, we also use $\chi^2$ statistics to estimate the model parameters in
equation (\ref{zL3p}) using the mean $L_{2\kev}$, $L_{2500}$ and $z$ of the
stacks of all RQQ's and of all quasars analysed in Section \ref{stackall},
together with the $L_{2\kev}$ errors estimated there. The number of RLQ stacks
is too few to get meaningful estimates for this many parameters.

\begin{figure}
\centering \fig[width=0.48\textwidth]{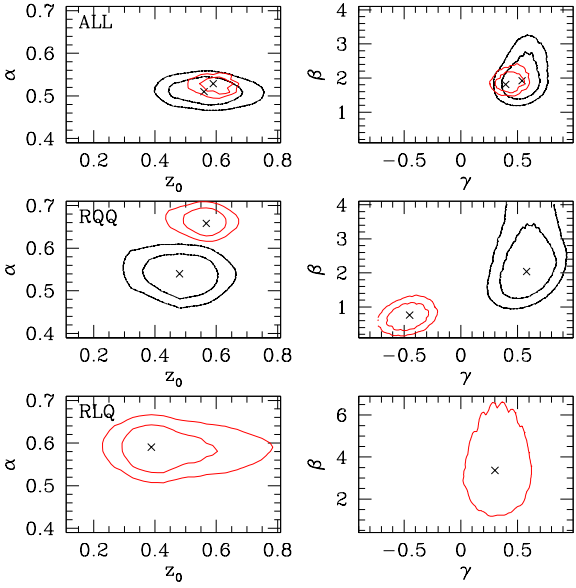} \caption{Maximum likelihood
estimates and confidence contours for model parameters in equations
(\ref{zL5p}). Results for the full QSO sample, RQQs and RLQs are shown in the
first, second and third rows respectively. Solid contours show 68.3\% and 90\%
confidence level contours obtained from the maximum likelihood analysis of
individual detections in Section \ref{depz}, while dashed contours give similar
contours based on $\chi^2$ fitting of the ``total'' stacks in Section
\ref{stackall}. Where only one contour is given it is the 68.3\% confidence
level contour.}
 \label{ML5p}
\end{figure}

We show maximum likelihood estimates and confidence contours with solid lines
for two pairs of model parameters in Fig. \ref{ML5p}. Results from $\chi^2$
fitting of the stacks are represented by dotted contours.  Model parameter
estimates from the two techniques are also listed in Tables \ref{table_ML} and
\ref{table_xi} respectively.  The minimum $\chi^2$ values listed in Table
\ref{table_xi} show that this more complex model for the redshift dependence
fits the mean $L_{2\kev}$ of our stacks significantly better than the simple
power-law model of equation (\ref{zL3p}). We note that the $\chi^2$ values for
RQQs are still significantly larger than the number of degrees of freedom in
fitting  equation (\ref{zL5p}). This is caused by several ``outliers'' with
small weighted errors, each contributing  $\Delta\chi^2>10$ in the fitting.
However, the inclusion or exclusion of these ``outliers'' does not change any
of our fitting results.

The top panels of Fig. \ref{ML5p} show that for our full set of quasars the
parameters of equation (\ref{zL5p}) are all well constrained. In addition, the
values obtained from the individual detections agree very well with those
obtained from the ``total'' stacks. Evolution is clearly detected in the sense
that the typical $L_{2\kev}$ associated with quasars of given $L_{2500}$
increases with redshift (both $\beta$ and $\gamma$ are significantly greater
than zero). This evolution is substantially stronger at low redshift than at
high ($\beta> \gamma$) and the transition redshift $z_0\sim 0.5$ is well
determined.  The $L_{2\kev} - L_{2500}$ relation is significantly shallower
than found above when fitting the ``total'' stacks without allowing for the
redshift dependence: $\alpha = 0.51$ here as opposed to $\alpha = 0.64$ in
equation (\ref{All}). This is a result of the strong selection-induced
correlation between $L_{2500}$ and redshift which is present in all our
samples.

For radio-loud quasars (bottom panels of Fig. \ref{ML5p}) the maximum
likelihood analysis gives parameters which, with one exception, are consistent
at about the 1-$\sigma$ level or better with those obtained for the sample as a
whole. The exception is the overall normalisation; as before we find that the
RLQs are systematically brighter in soft X-rays than the population as a whole
by about a factor of 2. The constraints on $\beta$ are relatively poor for
RLQs. The maximum likelihood value $\beta=3.4$ is substantially larger than
that for the sample as a whole ($\beta=1.8$) suggesting that radio quasars may
evolve more strongly than typical quasars at low $z$, but the difference is not
statistically significant. The ``total'' stacks for RLQs are fully consistent
with these parameters but do not contain enough information to constrain them
significantly.

The situation for radio-quiet quasars (middle panels of Fig. \ref{ML5p}) is
more complex. The parameter estimates we obtain from the ``total'' stacks are
very similar to those found for the quasar sample as a whole (and also to those
found for RLQs).  This is expected since about 90\% of AGN as a whole are radio
quiet. The parameter estimates we obtain from the individual RQQ detections
are, however, consistent neither with those obtained from the ``total'' RQQ
stacks nor with those found for the other two samples. Substantially weaker
evolution is indicated ($\beta$ and $\gamma$ are both much smaller) and, to
compensate, the slope found for the $L_{2\kev} - L_{2500}$ relation is steeper.
Clearly the two sets of results for the RQQs cannot both be correct, and it
seems suspicious that the RQQ behaviour should differ substantially from that
of the sample as a whole. For these reasons, and also because the ``total''
stacks use our dataset as a whole and extend out to $z\sim 4$ whereas there are
very few individual RQQ detections at $z>2$, we adopt the parameters obtained
from the RQQ ``total'' stacks as those best representing the population. We
present independent evidence below (Section \ref{stacks} and Fig.
\ref{simudata2}) that this is indeed the correct choice.

\subsubsection{The distribution of $L_{2\kev}$}\label{cdis}

In the last section, we assumed the distribution of $L_{2\kev}$ at given
$L_{2500}$ and $z$ to be log-normal with logarithmic dispersion $\sim0.40$
and we used maximum likelihood techniques to estimate its median value as a
function of $L_{2500}$ and $z$. In this section, we check the log-normal and
nearly constant scatter assumptions directly against our individual detection
data.

For simplicity, we here correct for redshift evolution using the results of
the last section, and we study how the distribution of redshift-corrected
$L_{2\kev}$ values depends on $L_{2500}$. Specifically, we correct the X-ray
luminosity of each individually detected quasar to redshift $z_0$ using
equation (\ref{zL5p}) with the maximum likelihood parameters of Table
\ref{table_ML}. For RQQs, we make this correction by using the $\chi^2$
parameter estimates listed in Table \ref{table_xi}. We denote the result by
$L_{2\kev,z_0}$. While we assume the redshift dependence deduced in subsection
\ref{depz} throughout the current section, we assumed the value of the scatter
deduced below when estimating this redshift dependence in Section \ref{depz}.
The results of the two sections are, in fact, consistent and were derived by
iteration.

We divide the individual detections into the same 7 $L_{2500}$ bins that we
used in Section \ref{stackall} for stacks.  The fraction of the quasars in each
$L_{2500}$ range which are individually detectable in the RASS at given
$L_{2\kev}$ depends on their redshifts and on the RASS detection sensitivity at
their positions. We use the the minimum luminosities $L_{\rm{min},j}$ of
Section \ref{depz} to calculate these fractions at each $L_{2\kev}$.  They can
then be used to correct the observed distribution of $L_{2\kev}$ for the
effects of the RASS flux limit and thus to estimate the unconstrained
distribution of $L_{2\kev}$ for quasars in the chosen $L_{2500}$ range. Each
detected quasar is simply weighted with a factor $1/F_{i}$, where $F_{i}$ is
defined by
\beq
F_{i}=\frac{N(L_{\rm{min}}<L_i)}{N_{tot}}.
\eeq
Here $N_{tot}$ is the {\it total} number of quasars in the $L_{2500}$ bin
(including non-detections and ``contaminants'') and $N(L_{\rm{min}}<L_i)$ is
the number of these for which the minimum detectable $L_{2\kev}$ is less than
the X-ray luminosity $L_i$ of the particular detected quasar under
consideration.

\begin{figure}\centering \fig[width=0.48\textwidth]{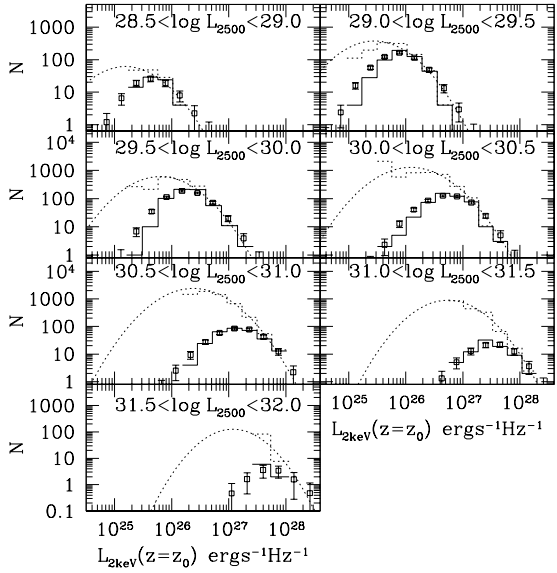}
\caption{The distribution of $L_{2\kev}$ for RQQs of given $L_{2500}$. Each
panel refers to a different range of $L_{2500}$ as indicated. Solid histograms
show the directly observed counts of detected quasars as a function of
$L_{2\kev}$. The dotted histograms show the result of weighting each quasar by
the inverse of the fraction of the (sub)sample in which its luminosity could
have been detected. This produces an estimate of the true X-ray luminosity
distribution in each $L_{2500}$ bin (see text for details). Maximum likelihood
log-normal fits to the distributions of individual detections are plotted as
dotted curves. The distributions of individual detections predicted from these
log-normal fits are shown as squares with error bars to denote the scatter
among individual Monte Carlo realizations of the luminosities predicted for the
quasars in each $L_{2500}$ range.} \label{simudata1}
\end{figure}

In Fig. \ref{simudata1}, we show the resulting luminosity distributions in
histogram form for the individually detected RQQs in each $L_{2500}$ bin,
together with the original un-weighted histograms of $L_{2\kev}$. After
correction, the sampled part of the luminosity distribution is well represented
by a log-normal in all cases. We do not show the histograms for RLQs, which are
similar but more noisy as a result of the limited number of individual
detections. Histograms for all detections, independent of radio properties, are
very similar to those shown here for the RQQs.

Based on these results we now {\it assume} the distribution of
evolution-corrected $L_{2\kev}$ at given $L_{2500}$ to be log-normal,
\begin{small}
 \beq\label{lognormal}
 P(L_{2\kev,z_0}|L_{2500}) =
      \frac{1}{\sqrt{2\pi}\sigma_{L_{2\kev}}}\exp\left[-\frac{\log
      (L_{2\kev,z_0}/\tilde{L}_{2\kev,z_0})^2}{2\sigma_{L_{2\kev}}^2}\right],
 \eeq
 \end{small}
and we use the maximum likelihood method of Section \ref{depz} to estimate the
two parameters $\tilde{L}_{2\kev,z_0}$ and $\sigma_{L_{2\kev}}$ (the median and
the logarithmic dispersion) for each $L_{2500}$ bin. The resulting best-fit
log-normal distributions are shown as dotted curves in each $L_{2500}$ panel
of Fig. \ref{simudata1}. They clearly agree very well with the directly
estimated $L_{2\kev, z_0}$ distributions.

\begin{figure}\centering \fig[width=0.48\textwidth]{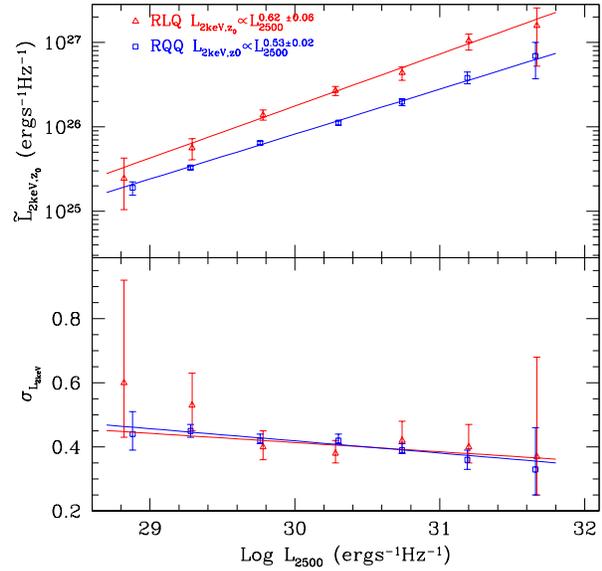}
\caption{The evolution-corrected X-ray luminosity distribution of quasars as a
function of $L_{2500}$. The median (upper panel) and the logarithmic scatter
(lower panel) of a log-normal fit to the evolution-corrected $L_{2\kev,z_0}$
distribution are shown separately for RLQs (open triangles) and for RQQs (open
squares). Error-bars denote 90\% confidence intervals. The $L_{2500}$ values
for the squares are shifted slightly for clarity.  These estimates are based on
individual detections only. The solid lines show least-squares error-weighted
fits to the points [equations (\ref{fitLxRL}) and (\ref{fitLxRQ})]. }
\label{fitL}
\end{figure}

Fig. \ref{fitL} shows these maximum likelihood estimates of the median and
scatter as functions of $L_{2500}$, with open triangles and open squares
denoting RLQs and RQQs respectively. Error bars indicate 90 percent confidence
intervals. The median value of $\log L_{2\kev,z_0}$ increases linearly with
$\log \bar{L}_{2500}$ both for RLQs and for RQQs. The relation for RLQs shows a
steeper slope and is systematically higher than that for RQQs Least squares
fits to these data give
\begin{small}
 \begin{eqnarray}\label{fitLxRL}
 \log \tilde{L}_{2\kev,z_0}=(0.62\pm0.06)(\log L_{2500}-30.5)+(26.55\pm0.04)
 \nonumber\\
\sigma_{L_{2\kev}}=(-0.03\pm0.04)(\log L_{2500}-30.5)+(0.40\pm0.02)
\end{eqnarray}
\end{small}
for RLQs and
\begin{small}
 \begin{eqnarray}\label{fitLxRQ}
 \log \tilde{L}_{2\kev,z_0}=(0.53\pm0.02)(\log L_{2500}-30.5)+(26.18\pm0.02)
 \nonumber\\
\sigma_{L_{2\kev}}=(-0.04\pm0.01)(\log L_{2500}-30.5)+(0.40\pm0.01)
\end{eqnarray}
\end{small}
for RQQs. Fitting to the full sample of QSOs gives
\begin{small}
 \begin{eqnarray}\label{fitLxAll}
 \log \tilde{L}_{2\kev,z_0}=(0.54\pm0.01)(\log L_{2500}-30.5)+(26.29\pm0.01)
 \nonumber\\
\sigma_{L_{2\kev}}=(-0.04\pm0.01)(\log L_{2500}-30.5)+(0.42\pm0.01)
\end{eqnarray}
\end{small}
The fits to RLQs and RQQs are plotted as solid straight lines in Fig.
\ref{fitL} and clearly represent the data very well. Furthermore they are
consistent both with our fits to the stacked data [equations (\ref{ARLQ}) and
(\ref{ARQQ}) and Fig. \ref{All_RQQ} and \ref{All_RLQ}; note that one must
correct for the offset between median and mean] and with the joint fits of
equation (\ref{zL5p}) for which parameters are given in Table \ref{table_xi}
and \ref{table_ML}. We detect no significant dependence of the scatter of the
$L_{2\kev}$ distribution on $L_{2500}$ and obtain a value consistent with
0.40 (as assumed in Section \ref{depz}) in all cases.

The parameter estimates of this section do not use any information about the
X-ray luminosities of individually undetected quasars.  Mean values of these
luminosities can, however, be measured by stacking. We now carry out such
stacking in order to check for consistency with the above results.

\subsection{Luminosities for stacks of undetected quasars}\label{stacks}

The results of the last section show that although the high X-ray luminosity
part of the distribution of $L_{2\kev, z_0}$ is well described by
a log-normal function, the shape of the distribution below the median
X-ray luminosity is not constrained by our individual detections. In this
section, we use stacking analysis to study the mean X-ray luminosity of
individually undetected sources, comparing our measurements with predictions
from the log-normal models of the last section.

The undetected sources were split into seven ranges of $L_{2500}$ and then
within each range they were separated into a series of bins according to
redshift. The ranges of $L_{2500}$ and redshift are the same as those of the
``total'' stacks  in Section \ref{stackall}. For the 51 stacks of RLQs and 118
stacks of RQQs, the numbers of detections with $L>7 $ (in either the 0.1--2.4
keV or the 0.5--2.0 keV band) are 49 and 117 respectively.

We use Monte-Carlo simulation to check whether the X-ray luminosities of these
stacks are consistent with the $L_{2\kev}$ distributions we have fit to the
individual detections alone. To do this, we randomly generate a value of
$L_{2\kev, z_0}$ for each real quasar using a log-normal distribution with
median and scatter given by the symbols in Fig. \ref{fitL}. After undoing the
evolution corrections using equation (\ref{zL5p}), we get a simulated
$L_{2\kev}$ for each quasar.  These Monte Carlo luminosities are then converted
into X-ray fluxes in the $0.1-2.4$\,keV band based on the quasar redshifts and
an assumed photon index $\Gamma$. As before, we take $\Gamma=1.9$ for RLQs and
$\Gamma=2.1$ for RQQs.  Using the RASS detection completeness function $C_{\rm
RASS}(f)$ described in Section \ref{comp}, we find the probability that each
source would be individually detected by the RASS.  Based on Monte-Carlo
sampling with this probability, we pick sources as ``individual RASS
detections''. Undetected quasars are then placed into the same redshift bins as
the real data and we simulate the stacking process using equations
(\ref{stack}) and (\ref{Lstack}). These procedures produce both individual and
stacked detections for a simulated sample with the chosen log-normal X-ray
luminosity distribution. We repeat this simulation 200 times and use the mean
and scatter among the 200 realizations to determine our model expectation and
its variance both for individual and for stacked detections.

\begin{figure}\centering \fig[width=0.48\textwidth]{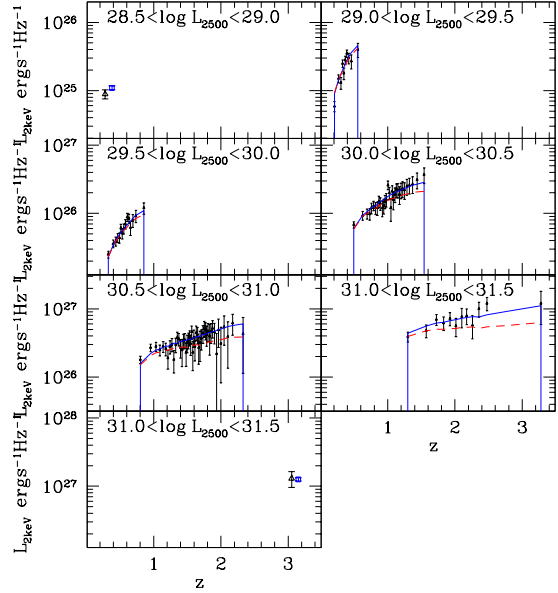}
\caption{The mean $L_{2\kev}$ of stacks of non-detected RQQs are compared to
predictions from the log-normal models fitted to individually detected RQQs and
to ``total'' stacks of RQQs. Within each $L_{2500}$ bin the mean $L_{2\kev}$
of each stack is plotted as a triangle with an error bar representing its
1-$\sigma$ observational uncertainty. The values predicted for these stacks by
the log-normal models are indicated by continuous solid lines for the model fit
to the ``total'' stacks, and by dashed lines for the model fit to the
individual detections. Clearly the former model is consistent with these
independent data, while the latter is not.} \label{simudata2}
\end{figure}

In Fig. \ref{simudata2} we compare the average X-ray luminosity of the
simulated stacks with the real data for RQQs.  The figures for RLQs and for the
quasar sample as a whole show equally good agreement and are not given here.
For each $L_{2500}$ bin, we show the mean $L_{2\kev}$ of stacks as function
of mean redshift. The observations are shown as triangles with error bars
giving the 1-$\sigma$ uncertainties in the measured flux. Model expectations
based on the parameter set obtained from the ``total'' stacks (Table
\ref{table_xi}) are shown by connected solid lines except in the lowest and
highest $L_{2500}$ bins which only have one stack. Note that the variance of
the model predictions is much smaller than the observational errors as shown
explicitly in the lowest and highest $L_{2500}$ panels. A similar prediction
based on the parameters found from our likelihood analysis of individual RQQ
detections (Table \ref{table_ML}) is shown by a dashed line. It is clear that
the prediction based on parameters from the ``total'' stacks is in excellent
agreement with these independent data, whereas that based on the maximum
likelihood analysis fits poorly at the higher redshifts. This provides further
support both for evolution parameters based on the ``total' stacks and for our
log-normal model for the X-ray luminosity distribution, reinforcing the fact
that the incompleteness-corrected $L_{2\kev}$ distributions of Fig.
\ref{simudata1} have log-normal shape on the bright side of the peak.

As a by-product from these Monte-Carlo simulations, Fig. \ref{simudata1} shows
model predictions for the distribution of $L_{2\kev,z_0}$ for individual
detections.  These are plotted as open squares with error bars representing the
scatter among the 200 simulations. These symbols should be compared with the
solid histograms which show number counts for individual detections in the real
data. Again there is good agreement, demonstrating the consistency of our
models and our completeness corrections.

\subsection{$L_{2\kev}$ versus [OIII] line luminosity}

In this section, we use [OIII]$\lambda$5007 line luminosity, $L_{[\rm{OIII}]}$,
instead of $L_{2500}$ to characterize the strength of quasar activity. As we
have mentioned before, only the 9103 lowest redshift quasars ($z<0.8$) have
measurements of [OIII]$\lambda$5007 line strength. Among them, there are 587
RLQs and 5880 RQQs. The distributions of $L_{[\rm{OIII}]}$ were shown as
histograms in the lower panels of Fig. \ref{RLRQ}. For RLQs, the number of
individual RASS detections is 190 and the number of random ``contaminants'' is
about 9. For RQQs, these two numbers are 1532 and 23.

\begin{figure}\centering \fig[width=0.48\textwidth]{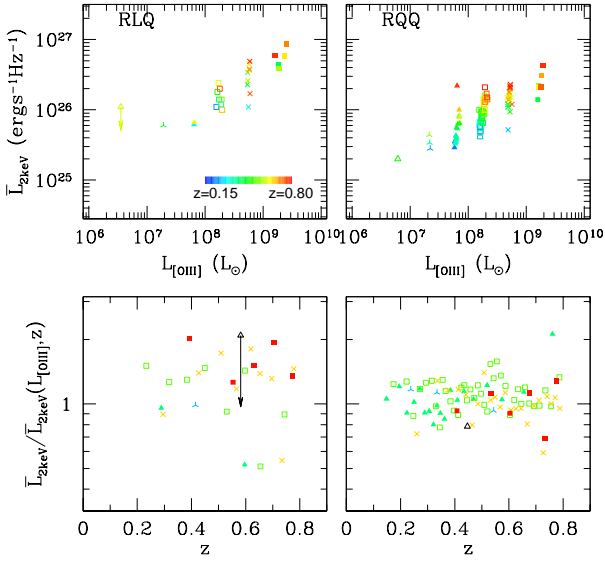}
\caption{$L_{2\kev}$ as a function of $L_{[\rm{OIII}]}$.  The left panels show
results for RLQs and the right panels for RQQs.  In each panel mean $L_{2\kev}$
is plotted as a function of mean $L_{[\rm{OIII}]}$ for stacks of all quasars in
a series of bins which are colour-coded according to their mean redshift. In
the bottom two panels, we divide  each stack's mean  $L_{2\kev}$ by the value
predicted for its mean $L_{[\rm{OIII}]}$ and $z$ by equation (\ref{L2kO3z}),
and we plot this ratio against mean redshift. Different symbol types and
colours here distinguish quasars in the different $L_{[\rm{OIII}]}$ bins. }
\label{L2k_O3}
\end{figure}

As for $L_{2500}$, we first stack the QSOs and study mean $L_{2\kev}$ as a
function of $L_{[\rm{OIII}]}$ and redshift. We divide objects into 6
$L_{[\rm{OIII}]}$ bins and stack them according to exposure time and rank in
redshift. We choose the total exposure time of a stack of RLQs to be $\sim 10$
ks and of a stack of RQQs to be $\sim 30$ ks.  This gives 26 stacks of RLQs and
90 stacks of RQQs. For the full sample, regardless of radio properties, this
gives 109 stacks. All stacks have RASS detections with $L>7$, except for the
stack of RLQs with the lowest $L_{[\rm{OIII}]}$ which contains only 3 QSOs.

In Fig. \ref{L2k_O3} (upper panels) we show the mean $L_{2\kev}$ of these
``total'' stacks as function of $L_{[\rm{OIII}]}$ for both RLQs and
RQQs. Symbols are colour-coded according to mean redshift.  At given
$L_{[\rm{OIII}]}$, RLQs again have larger $L_{2\kev}$ than RQQs. For RQQs,
there is a very clear trend with redshift in all the $L_{[\rm{OIII}]}$
bins. Higher redshift RQQs have higher mean $L_{2\kev}$ values. To quantify
this redshift dependence, we apply the $\chi^2$ fitting techniques of Section
\ref{stackall} to these stacks and the maximum likelihood methods of Section
\ref{depz} to the individual detections.

\begin{figure}
\centering \fig[width=0.48\textwidth]{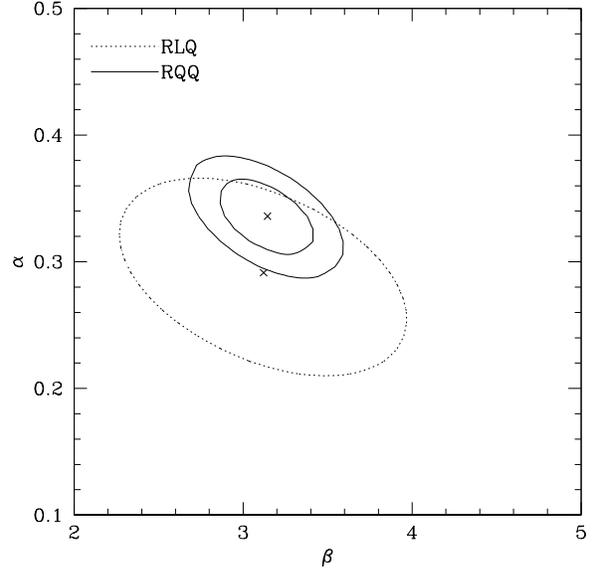}
\caption{Maximum likelihood estimates of the model parameters $\alpha$ and
 $\beta$ in equation (\ref{L2kO3z}) for RLQs and for RQQs. Solid lines show
 the 1- and $2-\sigma$ confidence contours for RQQs whereas the dotted
 line shows the 1-$\sigma$ confidence contour for RLQs.}
\label{O3z}
\end{figure}

\begin{figure}
\centering \fig[width=0.48\textwidth]{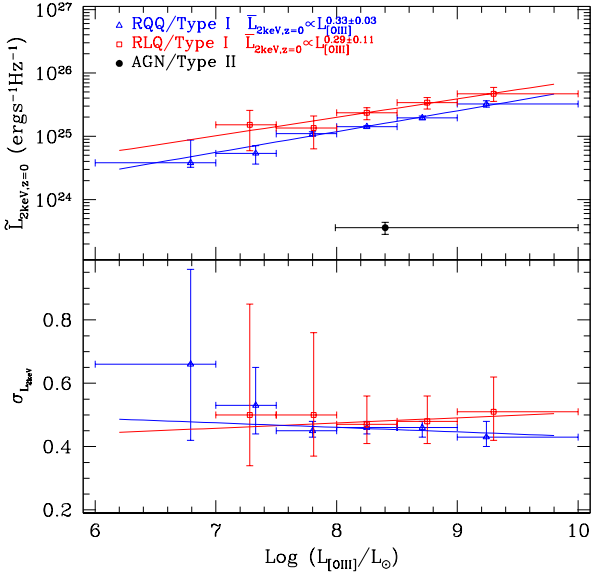}
\caption{The median and scatter of the evolution-corrected $\L_{2\kev,z=0}$
distribution at fixed $L_{[\rm{OIII}]}$. Horizontal error-bars indicate the
range of the $L_{[\rm{OIII}]}$ bins while vertical error-bars indicate 90
percent confidence intervals. Points are plotted at the mean $\log
L_{[\rm{OIII}]}$ of the quasars in each bin.  Triangles and squares show
results for RLQs and RQQs respectively, whereas the filled circle represents
the mean $L_{2\kev}$ value measured for type II AGN (see Section
\ref{typeII}). Solid lines are the least-squares fits of equations
(\ref{Lfit_O3RL}) and (\ref{Lfit_O3RQ}). } \label{fitL_O3}
\end{figure}

We again assume the distribution of $L_{2\kev}$ at given $L_{[\rm{OIII}]}$ and
$z$ to be log-normal with logarithmic dispersion $\sim0.45$ (see Fig.
\ref{fitL_O3}). We model the dependence of the median $L_{2\kev}$ on the
$L_{[\rm{OIII}]}$ and $z$ by
 \beq\label{L2kO3z}
 \log\tilde{L}_{2\kev}=\alpha(\log L_{[\rm{OIII}]}-8.5)+\beta\log(1+z)+\delta
 \eeq
Maximum likelihood estimates of model parameters are then  obtained from our
individual detections as in Section \ref{depz}. We show estimates of $\alpha$
and $\beta$ together with their confidence contours in Fig. \ref{O3z} for the
RLQ and RQQ subsamples. Parameter estimates for these two subsamples and for
the full sample of quasars with [OIII]$\lambda$5007 are also listed in Table
\ref{table_ML}. The values of $\alpha$ for the two samples differ at the
1-$\sigma$ level while the values of $\beta$ are almost identical.
Interestingly, the values of both parameters are quite different from those we
found above when characterising the optical luminosity of quasars by
$L_{2500}$. The optical luminosity dependence is weaker but the redshift
dependence is stronger.  In addition, the differences between RLQs and RQQs are
smaller and are not significant. The $1-\sigma$ confidence region for RLQ model
parameters is, however, quite large. In the bottom two panels of Fig.
\ref{L2k_O3}, we divide the mean $L_{2\kev}$ of the ``total'' stacks by the
values predicted by equation (\ref{L2kO3z}) (correcting for the difference
between mean and median) at their mean $L_{[\rm{OIII}]}$ and $z$. The resulting
ratios are plotted against mean $z$. Clearly the evolution inferred from our
analysis of individual detections does a very good job of removing systematic
redshift dependences from the ``total'' stacks also.

A $\chi^2$ analysis of the stacked data similar to that of Section
\ref{stackall} gives consistent conclusions. Fitting equation (\ref{L2kO3z}) to
the mean $L_{2\kev}$, $L_{[\rm{OIII}]}$ and $z$  values of the stacks shown in
Fig. \ref{L2k_O3} produces the parameter estimates listed in Table
\ref{table_xi}. These agree well with those derived from maximum likelihood
fits to the individual detections(Table \ref{table_ML}) and again the redshift
dependence is detected at very high significance. This is perhaps best seen by
comparing with the results of $\chi^2$ fitting to a simple, redshift-independent
power-law relation
\begin{equation}\label{L2kO3}
 \log\tilde{L}_{2\kev}=\alpha(\log L_{[\rm{OIII}]}-8.5)+\delta
\end{equation}
The results given in Table \ref{table_xi} show once more that substantially
larger values of $\alpha$ are inferred if the redshift dependence is ignored,
but, more importantly, the much larger minimum  $\chi^2$ values found for all
three samples when forcing $\beta=0$  demonstrate that a strong redshift
dependence really is required by  the data.

The power-law redshift dependence we assume here is similar to that of equation
(\ref{zL3p}) rather than to the more complex parametrisation of equation
(\ref{zL5p}) which we preferred above when analysing soft X-ray properties as a
function of $L_{2500}$. This choice reflects both the limited redshift range of
QSOs with measured [OIII]$\lambda$5007 ($z<0.8$), and the smaller samples
analysed here, which do not justify the inclusion of two additional model
parameters.

With the redshift dependence derived above, we ``correct'' all X-ray
luminosities to $z=0$ in order to study the dependence of $L_{2\kev,z=0}$ on
$L_{[\rm{OIII}]}$ in more detail. We divide the QSOs into six $L_{[\rm{OIII}]}$
bins with varying width and assume that the $L_{2\kev,z=0}$ values for QSOs in
each bin are distributed log-normally. We then use the maximum likelihood
method of Section \ref{cdis} to estimate the median and logarithmic dispersion
of this distribution. We show the results as functions of $L_{[\rm{OIII}]}$ in
Fig. \ref{fitL_O3}. A simple error-weighted least-squares fit gives
 \begin{small}
 \begin{eqnarray}\label{Lfit_O3RL}
 \log \tilde{L}_{2\kev,z=0} = (0.29\pm0.11)(\log L_{[\rm{OIII}]}-8.5)+(25.44\pm0.06)
 \nonumber \\
 \sigma_{L_{2\kev}}=(0.02\pm0.09)(\log L_{[\rm{OIII}]}-8.5)+(0.48\pm0.05)
\end{eqnarray}
\end{small}
for RLQs and
 \begin{small}
 \begin{eqnarray} \label{Lfit_O3RQ}
 \log \tilde{L}_{2\kev,z=0} = (0.33\pm0.03)(\log L_{[\rm{OIII}]}-8.5)+(25.24\pm0.02)
   \nonumber \\
 \sigma_{L_{2\kev}}=(-0.01\pm0.03)(\log L_{[\rm{OIII}]}-8.5)+(0.45\pm0.01)
\end{eqnarray}
\end{small}
for RQQs and
 \begin{small}
 \begin{eqnarray}\label{Lfit_O3}
 \log \tilde{L}_{2\kev,z=0} = (0.33\pm0.03)(\log L_{[\rm{OIII}]}-8.5)+(25.33\pm0.02)
  \nonumber\\
 \sigma_{L_{2\kev}}=(-0.01\pm0.02)(\log L_{[\rm{OIII}]}-8.5)+(0.43\pm0.01)
\end{eqnarray}
\end{small}
for the full sample.

All three samples have similar power law indices and a much shallower
$L_{2\kev,z=0} - L_{[\rm{OIII}]}$ relation than the $L_{2\kev,z_0} -
L_{2500}$ relation we found earlier.  The typical $L_{2\kev,z=0}$ of RLQs is
about 1.6 times larger than that of RQQs at given $L_{[\rm{OIII}]}$, slightly
smaller than the factor of 2 we found for the $L_{2500}$ case.  This is
because RLQs have higher $L_{[\rm{OIII}]}$ than RQQs with similar $L_{2500}$
(see Fig. \ref{RLRQ}). It is also interesting to note that the dispersion in
the $L_{2\kev}$ distribution is $\sim 0.45$ almost independent of luminosity,
slightly higher than that we found for the $L_{2500}$ case in Section
\ref{cdis}. Moreover, all the results here are consistent with our maximum
likelihood estimates of the model parameters in equation (\ref{L2kO3z})
(see Table \ref{table_ML}).

\subsection{Type II AGN}\label{typeII}

As mentioned earlier, the DR3 quasar catalogue includes only type I quasars.
For comparison, we now study the properties of a sample of type II AGN. This
is a set of 22623 narrow-line type II AGN selected from the SDSS main sample
of galaxies with $0.02<z<0.3$ (Kauffmann et al. 2003). The [OIII]$\lambda$5007
luminosity of each AGN was measured and used to characterize its nuclear
activity.

Applying our maximum likelihood X-ray detection procedure to these objects
gives only 211 individual detections with $L\geq 7$. These detections are
significantly biased towards objects with high $L_{[\rm{OIII}]}$. The number of
spurious detections is about 104, as estimated from a sample of random
positions with the same sky structure. The very low fraction of individual
detections and high fraction of contaminants suggest that the soft X-ray
emission of type II AGN is very weak. Thus, it is difficult to carry out the
same analysis as for type I quasars. Instead, we stack all the candidates to
derive their average X-ray fluxes. We first divide the type II AGN into ten
bins of $L_{[\rm{OIII}]}$, each containing the same number of objects. For each
$L_{[\rm{OIII}]}$ bin, we combine all the AGN into one stack. For these ten
stacks, we obtain only one detection with $L \geq 7$ (for the bin with the
highest $L_{[\rm{OIII}]}$). We calculate the weighted average of the X-ray
luminosity $L_{2\kev}$ from equation (\ref{Lstack}) for this $L_{[\rm{OIII}]}$
bin. The result is shown as the filled circle in Fig.  \ref{fitL_O3}. The mean
X-ray luminosity of type II AGN is about 75 times smaller than the median
luminosity of type I radio-quiet AGN with similar [OIII] line luminosity and
redshift. (The average redshift of the type II AGN in this bin is about 0.14.)
Since the mean X-ray luminosity of the type I AGN is about a factor of 2 larger
than their median X-ray luminosity, the effective shift in the $L_{2\kev}$
distribution between the two populations is probably about a factor of 150.

\section{Discussion and conclusions}

In this paper, we have studied the soft X-ray properties of quasars based on
the SDSS DR3 quasar catalogue and the RASS. The FIRST catalogue enabled us to
define subsamples of radio-loud and radio-quiet objects. We used both
individual and stack detections to investigate the X-ray properties of quasars
as a function of $L_{2500}$, of radio power, of $z$ and of $L_{[\rm{OIII}]}$.

By stacking all QSOs with similar optical luminosity and redshift, we have
shown that the average $L_{2\kev}$ of both RLQs and RQQs depends significantly
on redshift at given optical luminosity. At fixed UV continuum luminosity
$L_{2500}$ and at low redshift ($z<0.5$) the RLQ population may evolve more
strongly than the RQQs. At higher redshift, the X-ray brightening of RLQs and
RQQs of given $L_{2500}$ are both considerably slower.  If [OIII] line
luminosity $L_{[\rm{OIII}]}$ is substituted for $L_{2500}$, then RLQs and
RQQs show similar and strong evolution out to $z=0.8$. Assuming that the
redshift and optical luminosity dependences of the $L_{2\kev}$ distribution
separate, we correct all X-ray luminosities to a fiducial redshift and analyse
in more detail the shape of the distribution of the evolution-corrected X-ray
luminosity $\log L_{2\kev,z_0}$ at given optical luminosity. This distribution
is well approximated by a log-normal, at least for quasars more X-ray bright
than the median.  Adopting this model we find a tight but nonlinear relation
between $\tilde{L}_{2\kev}$ and optical luminosity ($L_{2500}$ or
$L_{[\rm{OIII}]}$). The dispersion in $\log L_{2\kev}$ is roughly $\sim 0.40$,
independent of optical luminosity. The typical X-ray luminosity of RLQs is
twice that of RQQs of the same $L_{2500}$, and 1.6 times that of RQQs of the
same $L_{[\rm{OIII}]}$. In addition, we find the average soft X-ray emission of
type II AGN to be more than 100 times weaker than that of radio-quiet type I
quasars of the same $L_{[\rm{OIII}]}$ and redshift.

Many of our results have been seen in previous studies. For example, the
nonlinear relation between $L_{2\kev}$ and $L_{2500}$ is consistent with that
of Vignali, Brandt \& Schneider (2003a); the result that RLQs have stronger
soft X-ray emission than RQQs is consistent with that of Brinkmann et al.
(2000); the result that soft X-ray emission from type II AGN is very weak is
consistent with that of Zakamska et al. (2004). There are also a number of new
results, for example, that $\alpha_{OX}$ depends not only on optical luminosity
but also on redshift. At given optical luminosity $L_{2500}$, the typical value
of $\alpha_{OX}$ decreases by about 0.3 from $z\approx0$ to $z\approx5$
[extrapolated from our best fits to equation (\ref{zL5p})] both for RLQs
and for RQQs. This is significant compared to earlier claims for the redshift
independence of $\alpha_{OX}$. For example, Strateva et al. (2005) quote
$\Delta\alpha_{OX}\approx0.03$ from $z\approx$0 to 5. The apparent change in
redshift dependence around $z\sim0.5$ is very interesting. We note that
rest-frame 2500\AA\, passes out of the SDSS spectral range at about this
redshift. We obtain the 2500\AA\, flux for such low redshift quasars by
extrapolation using the Vanden Berk et al. (2001) composite quasar spectrum, a
procedure which could plausibly introduce systematic biases into $L_{2500}$. To
check for such effects, we repeated our analysis replacing $L_{2500}$ by
$L_{3500}$, the continuum luminosity at rest-frame 3500\AA. This can be
estimated directly from the SDSS spectrum for all our quasars with $z<1.63$.
This substitution did not alter any of our conclusions about low-redshift
evolution either for RLQs or for RQQs. Another possible systematic might arise
from host-galaxy contributions to $L_{2500}$ in low-redshift ($z<0.5$) and
low-luminosity quasars. Strateva et al. (2005) addressed this issue in their
own study, concluding that the host-galaxy contribution does not exceed 20
percent.

The most detailed previous study of the relation between $L_{2\kev}$ and
$L_{2500}$ is that of Strateva et al. (2005), based on a complete sample of
228 radio-quiet non-BAL quasars more than 80\% of which were detected in
medium-deep pointed observations with ROSAT. The slope of $0.65\pm 0.02$ which
they quote for the $\log L_{2\kev}$--$\log L_{2500}$ relation is in excellent
agreement with the values we find for all our samples when we fit our
``total'' stacks without allowing for evolution [equation (\ref{ARQQ})].
However, after allowing for evolution, our analysis predicts a shallower
value, $0.53\pm0.02$, both from our individual detections and from our
``total'' stacks. This is a consequence of the significant evolution we
detect, together with the strong selection-induced correlation between optical
luminosity and redshift in our samples. Moreover, since $\log L_{2\kev}$ is
normally distributed with a scatter of about 0.40, the median $L_{2\kev}$ at
given $L_{2500}$ is smaller than the mean by a factor of about 1.7. The
excellent apparent agreement in Fig. \ref{All_RQQ} between the Strateva et
al. relation and our own RQQ relation is thus misleading. The Strateva et
al. result gives median $L_{2\kev}$ as a function of $L_{2500}$ since they fit
in logarithm space; on the other hand, our result gives the mean $L_{2\kev}$
of QSOs of given $L_{2500}$. Thus the relations should be off-set by a factor
of 1.7. Quasar variability could perhaps account for this discrepancy given
the longer X-ray exposure times in the Strateva et al. (2005) data. QSOs vary
on time-scales of $\sim 10^4$ seconds at X-ray wavelengths with typical
amplitudes of about 0.2 in $\log L_{2\kev}$ (e.g. Almaini et al. 2000;
Manners, Almaini \& Lawrence 2002). In our study such variability contributes
to the scatter in the $L_{2\kev}$ distribution at given optical luminosity,
since exposure times for our quasars are shorter than $\sim 10^4$s. Studies
based on longer X-ray exposures should find a distribution with a higher
median and lower scatter (but with the same mean). It is suggestive that
Strateva et al.  find a smaller scatter in $\log L_{2\kev}$ than we do, 0.29
(see Fig. 14 in their paper) rather than 0.40.

La Franca et al. (1995) and Yuan, Siber \& Brinkmann (1997) argued that
photometric error in the optical luminosities might cause an apparently weaker
than linear dependence of $L_{X}$ on $L_{O}$, even if the true relation is
linear.  This explanation is no longer tenable for data of the quality analysed
here [or indeed in Strateva et al. (2005)]. The relations we find really are
much weaker than linear, $\tilde{L}_{2\kev}\propto L_{2500}^{0.53}$(RQQs) and
$\tilde{L}_{2\kev}\propto  L_{[\rm{OIII}]}^{0.33}$. The difference between the
slopes for the two different optical measures of quasar activity is large and
interesting.

The most important assumption in our study which still remains to be fully
justified observationally is the log-normal form we adopt for the distribution
of $L_{2\kev}$ at given optical luminosity. As shown by Figures
\ref{simudata1} and \ref{simudata2}, all the observational data (e.g. the
fraction of individually detected objects, the distribution of their
$L_{2\kev}$, the mean $L_{2\kev}$ both for stacks of all sources and for
stacks of individually undetected sources) appear consistent with this
assumption, but they do not constrain the shape of the distribution at low
X-ray luminosities. Note, however, that the part of the distribution we can
characterize well already corresponds to more than 50\% of the sources and to
about 75\% of their total X-ray output. Our log-normal hypothesis will be more
thoroughly tested by future X-ray studies which will simultaneously achieve
wide area coverage and deep sensitivity limits, resulting in much greater
detection completeness than is possible with the RASS.

\section*{Acknowledgments}

The authors thank Iskra Strateva, Shude Mao and Xinwu Cao for helpful
discussion.  Shiyin Shen acknowledges the financial support from the exchange
program between the Chinese Academy of Sciences and Max-Planck-Gesellschaft.
This project is partly supported by NSFC10443002, 10403008, Shanghai Municipal
Science and Technology Commission No. 04dz\_05905.

Funding for the SDSS and SDSS-II has been provided by the Alfred P. Sloan
Foundation, the Participating Institutions, the National Science Foundation,
the U.S. Department of Energy, the National Aeronautics and Space
Administration, the Japanese Monbukagakusho, the Max Planck Society, and the
Higher Education Funding Council for England. The SDSS Web Site is
\url{http://www.sdss.org/}.

The SDSS is managed by the Astrophysical Research Consortium for the
Participating  Institutions. The Participating Institutions are the American
Museum of Natural History, Astrophysical Institute Potsdam, University of
Basel, Cambridge University, Case Western Reserve University, University of
Chicago, Drexel University, Fermilab, the Institute for Advanced Study, the
Japan Participation Group, Johns Hopkins University, the Joint Institute for
Nuclear Astrophysics, the Kavli Institute for Particle Astrophysics and
Cosmology, the Korean Scientist Group, the Chinese Academy of Sciences
(LAMOST), Los Alamos National Laboratory, the Max-Planck-Institute for
Astronomy (MPA), the Max-Planck-Institute for Astrophysics (MPIA), New Mexico
State University, Ohio State University, University of Pittsburgh, University
of Portsmouth, Princeton University, the United States Naval Observatory, and
the University of Washington.

{}

\newpage
\onecolumn

\begin{table}
\caption{The redshift range, $i$ band magnitude limit and number of objects for
each sample.  $N_{tot}$ is the total number of quasars. $N_x$ is the number of
these individually detected in RASS with $L>7$. $N_S$ is the number of stacks
used both when making ``total'' stacks of all quasars (section \ref{stackall})
and when stacking non-detections (section \ref{stacks}). $N_{S,A}$ is the
number of ``total'' stacks detected with $L>7$. The corresponding number for
stacks of non-detections is $N_{S,U}$. The samples labelled with [OIII] are for
quasars with [OIII]$\lambda$5007 measured. }
\begin{tabular}{lccccccc} \hline
Sample & redshift & magnitude & $N_{tot}$ & $N_x$  & $N_S$ & $N_{S,A}$ & $ N_{S,U}$\\
\hline
All &  $0.08<z<5.41$ & $i<20.5$ & 46420 & 3120 & 209 & 199 & 194 \\
RLQs & $0.10<z<5.31$ & $i<20.5$ & 3206 & 341 &  51 & 50 & 49\\
RQQs & $0.08<z<4.98$ & $i<19.1$ & 25705 & 2034 & 118 & 117 & 117 \\
$\rm{[OIII]}$      & $z<0.80$ & $i<20.5$ & 9103 & 2192 & 141 & 141 & 141 \\
$\rm{[OIII]}$  RLQs & $z<0.80$ & $i<20.5$ & 587 & 181 & 26 & 25  & 25 \\
$\rm{[OIII]}$  RQQs & $z<0.80$ & $i<19.1$ & 5880 & 1509 & 90 & 90 & 90 \\
\hline
\end{tabular}\label{sample}
\end{table}

\begin{table}
\caption{Estimates and 1-$\sigma$ confidence ranges for the model parameters in
equations (\ref{lfit}), (\ref{zL3p}), (\ref{zL5p}), (\ref{L2kO3z}) and
(\ref{L2kO3}) from $\chi^2$ fitting of the ``total'' stacks.  }
\begin{tabular}{lcccccccc} \hline
Case & equation  & $\alpha$ & $\beta$ &  $\gamma$ & $z_0$ & $\delta$  & $\chi_{min}^2$ \\
\hline
 ALL & (\ref{lfit}) & $0.64_{-0.01}^{+0.01}$ & $-$ & $-$  &  $-$ & $26.63_{-0.01}^{+0.01}$ & $306.0$
\\
 ALL & (\ref{zL3p}) & $0.52_{-0.02}^{+0.02}$ & $0.67_{-0.09}^{+0.09}$ & $-$  &  $-$ & $26.51_{-0.01}^{+0.02}$
  & $269.1$
\\
 ALL & (\ref{zL5p}) & $0.51_{-0.02}^{+0.02}$ & $1.89_{-0.33}^{+0.33}$ & $0.54_{-0.12}^{+0.08}$
 &  $0.56_{-0.04}^{+0.08}  $ & $26.55_{-0.01}^{+0.02}$ & $250.0$
 \\
 RQQ & (\ref{lfit}) & $0.68_{-0.01}^{+0.01}$ &  $-$ & $-$  &  $-$ & $26.53_{-0.01}^{+0.01}$ & $213.3$
 \\
 RQQ & (\ref{zL3p}) & $0.57_{-0.03}^{+0.02}$ &  $0.57_{-0.12}^{+0.16}$ & $-$  &  $-$ & $26.44_{-0.02}^{+0.02}$
 & $200.5$
\\
  RQQ & (\ref{zL5p}) & $0.54_{-0.04}^{+0.02}$ & $2.04_{-0.52}^{+0.72}$ & $0.54_{-0.14}^{+0.18}$
  &  $0.48_{-0.06}^{+0.08}  $ & $26.44_{-0.03}^{+0.03}$ & $190.7$
  \\
RLQ & (\ref{lfit}) & $0.65_{-0.03}^{+0.02}$ &  $-$ & $-$  &  $-$ &
$26.91_{-0.01}^{+0.02}$ & $35.2$
 \\
 RLQ & (\ref{zL3p}) & $0.63_{-0.06}^{+0.05}$ &  $0.12_{-0.29}^{+0.28}$ & $-$  &  $-$ & $26.89_{-0.05}^{+0.05}$
 & $35.0$
\\
$\rm{[OIII]}$ ALL & (\ref{L2kO3}) & $0.46_{-0.02}^{+0.02}$ & $-$ & $-$ & $-$ &
$26.18_{-0.01}^{+0.01}$ & 209.2
\\
$\rm{[OIII]}$ ALL & (\ref{L2kO3z})&  $0.33_{-0.02}^{+0.02}$ &
$2.68_{-0.24}^{+0.18}$ & $-$ & $-$ & $25.68_{-0.03}^{+0.04}$ & 97.2
\\
$\rm{[OIII]}$ RLQ & (\ref{L2kO3}) & $0.55_{-0.07}^{+0.07}$ & $-$ & $-$ & $-$ &
$26.36_{-0.04}^{+0.04}$ & 13.5
\\
$\rm{[OIII]}$ RLQ  & (\ref{L2kO3z}) & $0.45_{-0.08}^{+0.08}$ &
$2.38_{-1.03}^{+0.96}$ & $-$ & $-$ & $25.91_{-0.17}^{+0.19}$ & 8.7
\\
$\rm{[OIII]}$ RQQ & (\ref{L2kO3}) & $0.49_{-0.02}^{+0.03}$ & $-$ & $-$ & $-$ &
$26.10_{-0.01}^{+0.01}$ & 141.6
\\
$\rm{[OIII]}$ RQQ  & (\ref{L2kO3z}) & $0.32_{-0.03}^{+0.02}$ &
$3.26_{-0.26}^{+0.29}$ & $-$ & $-$ & $25.49_{-0.05}^{+0.05}$ & 39.1
\\
 \hline
\end{tabular}\label{table_xi}
\end{table}

\begin{table}
\caption{Maximum likelihood estimates and 1-$\sigma$ confidence ranges for the
model parameters in equations  (\ref{zL5p}) and (\ref{L2kO3z}). }
\begin{tabular}{lcccccc} \hline
Case & equation  & $\alpha$ & $\beta$ &  $\gamma$ & $z_0$ & $\delta$  \\
\hline
 RLQ & (\ref{zL5p}) & $0.58_{-0.03}^{+0.03}$ & $3.39_{-1.45}^{+1.75}$ & $0.47_{-0.20}^{+0.16}$
  & $0.39_{-0.06}^{+0.06}$ & $26.66_{-0.08}^{+0.14}$ \\
 RQQ & (\ref{zL5p}) & $0.66_{-0.02}^{+0.01}$ & $0.71_{-0.26}^{+0.24}$ & $-0.48_{-0.11}^{+0.12}$
  &  $0.55_{-0.02}^{+0.04}  $ & $26.28_{-0.02}^{+0.02}$  \\
 ALL & (\ref{zL5p}) & $0.53_{-0.01}^{+0.02}$ & $1.81_{-0.14}^{+0.28}$ & $0.40_{-0.06}^{+0.10}$
  &  $0.59_{-0.01}^{+0.04}  $ & $26.30_{-0.02}^{+0.01}$  \\
$\rm{[OIII]}$ RLQ & (\ref{L2kO3z})  & $0.29_{-0.05}^{+0.04}$  &
$3.12_{-0.54}^{+0.54}$ &     $-$    &  $-$     & $25.44_{-0.02}^{+0.02}$ \\
$\rm{[OIII]}$ RQQ & (\ref{L2kO3z})  & $0.34_{-0.01}^{+0.01}$  &
$3.14_{-0.16}^{+0.18}$ &     $-$    &  $-$   & $25.25_{-0.01}^{+0.01}$ \\
$\rm{[OIII]}$ ALL & (\ref{L2kO3z})  & $0.31_{-0.01}^{+0.01}$  &
$3.22_{-0.14}^{+0.12}$ &     $-$    &  $-$   & $25.31_{-0.01}^{+0.01}$ \\
 \\
\hline
\end{tabular}\label{table_ML}
\end{table}

\twocolumn

\end{document}